\documentstyle[amssymb,aps]{revtex}

\begin{document}
\title{DNA folding: structural and mechanical properties of the two-angle model for
chromatin}
\author{Helmut Schiessel$^{\text{1,2}}$\thanks{%
Present address: Max-Planck-Institute for Polymer Research, Theory Group,
POBox 3148, 55021 Mainz, Germany}, William M. Gelbart$^{\text{2}}$, and
Robijn Bruinsma$^{\text{1}}$\thanks{%
Present address: Instituut-Lorentz for Theoretical Physics, Universiteit
Leiden, Postbus 9506, 2300 RA Leiden, The Netherlands}}
\address{$^{\text{1}}$Departments of Physics and $^{\text{2}}$Chemistry and\\
Biochemistry, University of California, Los Angeles, Los Angeles, CA 90095\\
\bigskip }
\maketitle

\begin{abstract}
\setlength{\baselineskip}{12pt}ABSTRACT We present a theoretical analysis of
the structural and mechanical properties of the 30-nm chromatin fiber. Our
study is based on the two-angle model introduced by Woodcock et al.
(Woodcock, C. L., S. A. Grigoryev, R. A. Horowitz, and N. Whitaker. 1993. 
{\it Proc. Natl. Acad. Sci. USA. }90:9021-9025) that describes the chromatin
fiber geometry in terms of the entry-exit angle of the nucleosomal DNA and
the rotational setting of the neighboring nucleosomes with respect to each
other. We explore analytically the different structures that arise from this
building principle, and demonstrate that the geometry with the highest
density is close to the one found in native chromatin fibers under
physiological conditions. On the basis of this model we calculate mechanical
properties of the fiber under stretching. We obtain expressions for the
stress-strain characteristics which show good agreement with the results of
recent stretching experiments (Cui, Y., and C. Bustamante. 2000. {\it Proc.
Natl. Acad. Sci. USA. }97: 127-132) and computer simulations (Katritch, V.,
C. Bustamante, and W. K. Olson. 2000. {\it J. Mol. Biol. }295:29-40), and
which provide simple physical insights into correlations between the
structural and elastic properties of chromatin.
\end{abstract}

\setlength{\baselineskip}{12pt}\bigskip 

Running title: DNA folding

Key words: {\it chromatin, 30-nm fiber, nucleosomes, fiber stretching, DNA
bending\newpage }

\section{Introduction}

Recently there has been considerable progress both in the visualization
(Bednar et al., 1998) and micromanipulation (Cui and Bustamante, 2000) of
chromatin fibers. These results constitute an important step towards the
understanding of DNA ''folding'', i.e., the problem of how plant and animal
genomes organize themselves into volumes whose linear dimensions are many
orders of magnitude smaller than their contour lengths. For instance, human
DNA is billions of base pairs (bp) long (about a meter), this length of
highly-charged (about one fundamental charge per two Angstroms) and
hard-to-bend (persistence length of $50nm$) linear polymer must be condensed
into chromosomes that fit into cell nuclei whose characteristic size is a 
{\it micron}.

An important part of the condensation process is the complexation of DNA
with oppositely-charged globular protein (histone) aggregates that have the
shape of squat cylinders. These aggregates are octameric complexes
consisting of pairs of the four core histones H2A, H2B, H3, and H4. A DNA
stretch of 147 bp's is wrapped in a $1$-$\,$and-$\,3/4$ left-handed
superhelical turn around the histone octamer and is connected via a stretch
of ''linker'' DNA to the next such protein spool. Each protein aggregate
together with its wrapped DNA comprises a {\it nucleosome core particle}
(cf. Fig. 1) with a radius of about 5nm and a height of about 6nm; with its
linker DNA it is the fundamental chromatin repeating unit. It carries a
large negative electrostatic charge (Khrapunov et al., 1997; Raspaud et al.,
1999). Whereas the structure of the core particle has been resolved up to
high atomic resolution (Luger et al., 1997), there is still considerable
controversy about the nature of the higher-order structures to which they
give rise. When stretched, the string of DNA/histone complexes has the
appearance of ''beads-on-a-string''. This basic structure can be seen
clearly when chromatin is exposed to very low salt concentrations, and is
known as the 10-{\it nm fiber} (Thoma et al., 1979), since the diameter of
the core particle is $10nm$. With increasing salt concentration, i.e.,
heading towards physiological conditions ($100mM$), this fiber appears to
thicken, attaining a diameter of $30nm$ (Widom, 1986). The absence of the
extra ''linker histones'' (H1 or H5) leads to more open structures (Thoma et
al., 1979) so it is surmised that the linker histones act near the
entry-exit point of the DNA (cf. Fig. 1); they carry an overall positive
charge and seem to bind the two strands together leading to a stem formation
(Bednar et al., 1998). Increasing the salt-concentration is expected to
decrease the entry-exit angle of the stem as it reduces the electrostatic
repulsion between the two strands.

Longstanding controversy (van Holde, 1989; Widom, 1989; van Holde and
Zlatanova, 1995, 1996) surrounds the structure of this 30{\it -nm fiber},
for which there are mainly two competing classes of models: the solenoid
models (Finch and Klug, 1976; Thoma et al., 1979; Widom and Klug, 1985); and
the zig-zag or crossed-linker models (Woodcock et al., 1993; Horowitz et
al., 1994; Leuba et al., 1994; Bednar et al., 1998). In the solenoid model
(Fig. 2a) it is assumed that the chain of nucleosomes forms a helical
structure with the axis of the core particles being perpendicular to the
solenoid axis (the axis of an octamer corresponds to the axis of the
superhelical path of the DNA that wraps around it). The DNA entry-exit side
faces inward towards the axis of the solenoid. The linker DNA (shown as a
dashed curve at the top of Fig. 2a) is required to be bent in order to
connect neighboring nucleosomes in the solenoid. The other class of models
posits {\it straight} linkers that connect nucleosomes located on {\it %
opposite} sides of the fiber. This results in a three-dimensional
zig-zag-like pattern of the linker (Fig. 2b).

Images obtained by electron cryomicroscopy should in principle be able to
distinguish between the structural features proposed by the different models
mentioned above (Bednar et al., 1998). The micrographs show a zig-zag motif
at lower salt concentrations and they indicate that the chromatin fiber
becomes more and more compact when the ionic strength is raised towards the
physiological value. However, for these denser fibers it is still not
possible to detect the exact linker geometry\footnote{%
Experiments on dinucleosomes (two nucleosomes connected by one linker) have
been performed to check if the nucleosomes ''collapse'' upon an increase in
ionic strength. A collapse would only occur if the linker bends, and an
observation of this phenomenon would support the solenoid model. The
experiments by Yao et al. (Yao et al., 1990) as well as more recent
experiments by Butler and Thomas (Butler and Thomas, 1998) indeed reported a
bending of the linkers but do not agree with experiments by Bednar et al.
(Bednar et al., 1995) and by others that did not find any evidence for a
collapse. Critical discussions of these and other experiments on
dinucleosomes are available (van Holde and Zlatonova, 1996; Widom, 1998).}.

An important experimental achievement was the stretching of a single
chromatin fiber via micromanipulation (Cui and Bustamante, 2000). The
''force-extension'' measurements show a rich behavior of the mechanical
properties as a function of the ionic strength. At low ionic strength ($5mM$
NaCl) the force-extension curves are reversible as long as the tension does
not exceed $20pN$. For higher tension levels ($\gtrsim 20pN$) there are
irreversible changes that lead to an increase of the fiber length, probably
due to the loss of linker histones and/or histone octamers. At high ionic
strength ($40mM$ and $150mM$ NaCl) a $5pN$-plateau in the force extension
curve was found\footnote{%
Marko and Siggia (1997) had in fact proposed an elastic model which
predicted a coexistence regime in the force-extension curve, with
nucleosomes ''evaporating'' from the fiber at higher force levels of the
order of $2pN$ which would lead to extensive irreversibility in the
force-extension curve. Although irreversibility is encountered at high force
levels, as mentioned, the $5pN$ plateau is reversible indicating that there
was no nucleosomal loss.}. The authors interpreted their results as
indicating a coexistence between ''swollen'' and ''condensed'' parts of the
fiber. In order to reproduce the force-extension curves, Katritch et al.
performed Monte Carlo simulations (Katritch et al., 2000) based on a
geometrical ''two-angle'' model introduced by Woodcock et al. for the 30-nm
fiber (Woodcock et al., 1993) combined with the worm-like chain (''WLC'')
Hamiltonian for the linkers. The WLC is widely used for predicting the
mechanical properties of {\it naked} DNA. The low-salt behavior could be
reproduced for several sets of angles and bond lengths of the model,
demonstrating both that it is a reasonable model and that it is not possible
to deduce a unique structure from the measured response of the fiber under
stretching. Katritch et al. found that an internucleosomal attraction of
roughly $3kT$ might explain the experimentally observed plateau in the
force-extension profile. The biological importance of these results lies in
the fact that significant changes can be achieved in the degree of chromatin
condensation with only modest levels of external stress. The fact that
chromatin at physiological salt concentrations apparently can exist in two
alternative forms that interconvert under low levels of stress is
particularly interesting.

The success of the model motivated the present study to provide an {\it %
analytical} framework for understanding the geometrical and mechanical
properties of the 30-nm fiber based on the two-angle model. Our first main
result is the derivation of a general structural phase diagram of the
chromatin fiber as a function of the two basic angles $\theta $ and $\phi $
determined by the nucleosome properties and the linker length $b$ (see
below). The various solenoidal, zig-zag and crossed-linker structures -- all
of which are assumed to have straight linkers -- appear as ''points'' in
this phase diagram (see Fig. 4) . We find that, within the two-angle model,
the position of chromatin fibers at physiological conditions (the ''native''
fibers) in the phase diagram is surprisingly close to the point in the
diagram with the {\it highest density} and the {\it maximal accessibility},
consistent with excluded-volume restrictions. Changes in bond angles induced
by physico-chemical changes in the environment lead to predictable changes
of the fiber away from this optimal point towards more open structures.

Our second main result is that we can obtain approximate analytical results
for the bending stiffness of the two-angle model -- and hence for the
persistence length -- and for the force-extension curve. We find ({\it i})
that the persistence length of the fiber should be comparable or less than
that of naked DNA, for a wide range of $\theta $- and $\phi $-values and (%
{\it ii}) that the stretching modulus should be so low that there is no
longer a pronounced difference between ''soft'' entropic elasticity (for low
forces) and ''hard'' entropic elasticity (for high forces), in marked
contrast with the case of naked DNA. Using the estimated values of $\theta $%
, $\phi $ and $b$ the predicted force-extension curves ({\it with no fitting
parameter}) are in good agreement with the data found for the stretching of
chromatin fibers (Cui and Bustamante, 2000).

The implication of our results is that a swollen 30-nm fiber should be very 
{\it soft} in terms of its elastic properties, over a wide range of values
of the angle parameter $\theta $ and $\phi $, a very reasonable ''design
feature'' in terms of its biological role. This swollen state competes with
a more rigid condensed state that appears, as a function of bond angle $%
\theta $, when we allow for (weak) attractive forces between nucleosomes.
The physical properties of the condensed state are beyond the scope of the
current paper, but the condensed fiber is expected to be significantly
stiffer that the swollen fiber. In general, our results appear to indicate
that the ''engineering design'' of the 30-nm fiber combines high compaction
levels with high structural accessibility and flexibility. Independent of
the question whether the swollen or the condensed state is realized, modest
changes in the control parameter $\pi -\theta $ (the nucleosome entry-exit
angle) produce large structural changes.

The paper is organized as follows. In the Section 2 we derive the
geometrical properties of the two-angle model and present the general
diagram of states. In Section 3 we apply our results of the two-angle model
to interpret the structure of the 30-nm chromatin fiber in terms of simple
optimization principles. Section 4 derives the elastic properties of the
two-angle model and gives the bending stiffness and the force-extension
relation. In the concluding section we summarize our results and discuss
alternative models.

\section{The two-angle model: folded structures}

\subsection{General relationships}

To address the folding problem of DNA at the level of the 30-nm fiber we
need a mathematical description for the different possible folding pathways.
At the simplest level, it is assumed that the geometric structure of the
30-nm fiber can be obtained from the intrinsic, single-nucleosome structure.
The specific roles of linker elastic energy, nucleosome-nucleosome
interaction, preferred binding sites, H1 involvement, etc. will be treated
afterwards as ''corrections'' to this basic model. To see how
single-nucleosome properties can control the fiber geometry, consider the
fact that DNA is wrapped a non-integral number of turns around the
nucleosome, e.g., 1-and-$3/4$ times (147 bp's) in the case of no H1. This
implies that the incoming and outgoing linker chains make an angle $\theta $
with respect to each other --- the entry-exit angle $\pi -\theta $ is
nonzero. In the presence of the histone H1 (or H5) the in- and outcoming
linker are in close contact forming a stem before they diverge (Bednar et
al., 1998). While the precise value of the resulting exit-angle depends on
salt concentration, degree of acetylation of the histones, etc., we may
nevertheless assume $\theta $ to be a quantity {\it that is determined
purely at the single-nucleosome level}. Next, we define the rotational
(dihedral) angle $\phi $ between the axis of neighboring histone octamers
along the necklace (see Fig. 3). Because nucleosomes are rotationally
positioned along the DNA, i.e., adsorption of DNA always begins with the
minor groove turned in towards the first histone binding site, the angle $%
\phi $ is a {\it periodic} function of the linker length $b$, with the $10bp$
repeat length of the helical twist of DNA as the period. There is
experimental evidence that the linker length shows a preferential
quantization involving a set of values that are related by integral
multiples of this helical twist (Widom, 1992), i.e., there is a preferred
value of $\phi $. (Note that the ''linker length'' $b$ is strictly speaking
defined here as the distance between two neighboring nucleosomes, cf. Fig.
3.)

If we treat the pair of angles $\left( \theta ,\phi \right) $, together with
the linker length $b$, as given physical properties (even though in vivo
they are likely under biochemical control), then the geometrical structure
of the necklace is determined entirely by $\theta $, $\phi $ and $b$. The
model only describes linker geometry and does not account for excluded
volume effects and other forms of nucleosome-nucleosome interaction; it
assumes that the core particles are pointlike ($a=0$) and that they are
located at the joints of the linkers. The model also assumes that the
linkers are {\it straight}. It is under dispute whether this last condition
holds for the 30-nm fiber at higher salt concentrations, and we will return
to this issue later. The $\left( \theta ,\phi \right) $-model is similar to
the freely rotating chain model encountered in polymer physics literature
(see, for instance, Doi and Edwards, 1986). The main difference is that in
the present case there is no free rotation around the linker and so torsion
is transmitted (see also Plewa and Witten, 2000).

As shown in Appendix A it is now possible to construct a spiral of radius $R$
and pitch angle $\gamma $ such that the nucleosomes - but not necessarily
the linker chain - are located on this spiral. The nucleosomes are placed
along the spiral in such a way that successive nucleosomes have a fixed
(Euclidean) distance $b$ from one another. From straightforward geometrical
considerations we can derive analytical expressions that relate pitch angle $%
\gamma $ and radius $R$ of the solenoid to the pair of angles $\theta $, $%
\phi $ and linker length $b$. Specifically the linker length $b$ can be
expressed as a function of $\gamma $, $R$ and $s_{0}$ (defined as the
vertical distance between successive ''nucleosomes'' along the helical
axis), $b=b\left( \gamma ,R,s_{0}\right) $, as given by Eq. \ref{s0}. The
corresponding relationships for the angles $\theta $ and $\phi $, $\theta
=\theta \left( \gamma ,R,s_{0}\right) $ and $\phi =\phi \left( \gamma
,R,s_{0}\right) $, are given by Eq. \ref{cost} and Eq. \ref{cosf}. Using
these relations, we can construct a catalog of structures.

\subsection{Planar structures}

If either one of the angles $\theta $ or $\phi $ assumes the value $0$ or $%
\pi $, then the resulting structure is {\it planar}, and calculation of the
associated geometrical properties is straightforward. Let us start with the
case $\phi =0$. If we also have $\theta =0$ the fiber forms a straight line
(see structure ''1'' in Fig. 4). For small non-vanishing $\theta $ the
structure forms a {\it circle} of radius $R\simeq b/\theta $. For the
special case $\theta =2\pi /n$, with $n$ an integer, the ring contains $n$
monomers before it repeats itself and we obtain a {\it regular polygon} (see
''2''). The special case is $\theta =\pi /2$ corresponds to the square
\'{(}''3''). With increasing $\theta $ the radius of the circle shrinks and
approaches asymptotically the value $b/2$. For $\theta =\pi \left(
n-1\right) /n$ with $n$ being an odd integer one encounters a series of
closed {\it star-like polygons} with $n$ tips. In particular, $n=3$
corresponds to the equilateral triangle (''4''), $n=5$ to the regular
pentagram (''5''), etc.

Next we consider the case $\phi =\pi $ and $\theta $ arbitrary. This case
corresponds to 2D {\it zig-zag-like structures}, as shown by ''6'' and ''7''
at the top of Fig. 4. The length of a fiber consisting of $N$ monomers is
given by 
\begin{equation}
L=b\cos \left( \theta /2\right) N  \label{L}
\end{equation}
and the diameter is given by $D=b\sin \left( \theta /2\right) $. Note that
the length of the fiber increases with decreasing $\theta $.

To complete our discussion of planar structures we mention the remaining
cases: $\theta =0$ with an arbitrary value of $\phi $ leads to the straight
line mentioned earlier (''1''); $\theta =\pi $ and arbitrary $\phi $
corresponds to linkers that go back and forth between two positions (''8'').

\subsection{Three-dimensional fibers}

({\it i}) {\it Solenoids}: For small angles, $\theta \ll 1$ and $\phi \ll 1$%
, we find structures that resemble solenoids where the linkers themselves
follow closely a helical path (see ''9'' in Fig. 4). For these structures
one has $\alpha s_{0}/R\ll 1$ --- where $\alpha =\cot \gamma $, with $\gamma 
$ the pitch angle{\it .} To lowest order in $\alpha s_{0}/R$ we find $%
b\simeq s_{0}\sqrt{1+\alpha ^{2}}$ (cf. Eq. \ref{s0}), $\theta \simeq \alpha
^{2}b/\left( R\left( 1+\alpha ^{2}\right) \right) $ (cf. Eq. \ref{cost}),
and $\phi \simeq \theta /\alpha $ (cf. Eq. \ref{cosf}). From this we can
infer several geometrical properties of the fiber as a function of $b$, $%
\theta $ and $\phi $ summarized in Table I. $R$ denotes the radius of the
fiber, $L$ is the length of a fiber consisting of $N+1$ monomers, $\lambda $
is its line density $N/L$ and $\rho $ is the 3D density given by $\rho
=\lambda /\pi R^{2}$, assuming a hexagonal array.

Other geometrical information can be obtained easily. For instance, the
vertical distance $d$ between two loops follows from $L$ in Table I by
setting $N=2\pi /\theta $ (the number of monomers per turn): 
\begin{equation}
d\simeq \frac{2\pi \phi b}{\theta \sqrt{\phi ^{2}+\theta ^{2}}}  \label{d}
\end{equation}
Furthermore, the pitch angle $\gamma $ is given by 
\begin{equation}
\cot \gamma \simeq \frac{\theta }{\phi }  \label{alpha2}
\end{equation}
$\gamma $ decreases monotonically as the ratio of the angles, $\theta /\phi $%
, increases. For $\phi \ll \theta $ one finds $\gamma \simeq \phi /\theta $.
In this regime one has a very dense spiral with $d\ll R$. In the opposite
limit $\phi \gg \theta $ the pitch angle is very large, namely $\gamma
\simeq \pi /2-\theta /\phi $ and the solenoid has a very open structure with 
$d\gg R$.\footnote{%
We note that such an open structure could in principle collapse into a very
dense fiber like the solenoidal model proposed by Klug (cf. Fig. 2a) if we
would allow the linkers to bend. As mentioned already before it is still a
matter of controversy if such linker bending takes place in chromatin.We
will stick in this study to the assumption of straight linkers.}$^{,}$%
\footnote{%
We mention that in the limit $\phi \rightarrow 0$ we recover the planar
circle with radius $R\simeq b/\theta $, cf. Table I.}

({\it ii}) {\it Fibers with crossed linkers}: Consider structures where $%
\phi $ is still small but where the entry-exit angle $\theta $ is large,
i.e. $\pi -\theta \ll \pi $. We discussed in the previous section that for $%
\phi =0$ one encounters star-shape polygons that are closed for $\theta =\pi
\left( n-1\right) /n$ with $n$ odd. For {\it non}-vanishing $\phi \ll 1$ the
star-shaped polygons open up in an accordion-like manner. This leads to a
three-dimensional fiber with crossed linkers -- see ''10''. It follows from
Eqs. \ref{s0} and \ref{cosf} that $s_{0}^{2}\simeq \phi ^{2}\left(
4R^{2}-b^{2}\right) /4$ for $\phi \ll 1$. Using this result as well as Eq. 
\ref{cost} $R$, $L$, $\lambda $ and $\rho $ can be expressed as a function
of $b$, $\theta $ and $\phi $, cf. Table I.

Assume now that $\theta _{n}=\pi \left( n-1\right) /n$ so that the
projection of the fiber is a closed polygon (this is only strictly true for $%
\phi =0$ but it is still a good approximation for $\phi \ll 1$). We can
calculate for this case the spacial distance $d$ between nucleosome $i$ and $%
i+n$: 
\begin{equation}
d\simeq \frac{n\phi b}{2}\cot \left( \theta /2\right) \simeq \frac{\pi \phi b%
}{4}\left( 1+\frac{\pi ^{2}}{12n^{2}}\right)  \label{d2}
\end{equation}

({\it iii}) {\it Twisted zig-zag structures}: Finally, we discuss structures
with a rotational angle $\phi $ close to $\pi $, say $\phi =\pi -\delta $
with $\delta \ll 1$. For $\delta =0$ we recover the 2D zig-zag structure
discussed earlier (''6'' and ''7''). Small non-vanishing values of $\delta $
lead to {\it twisted }zig-zag structures{\it \ }-- see ''11''. In this case
monomer $i+1$ is located nearly opposite to the $i$th monomer, but slightly
twisted by an angle $\delta $. Monomer $i+2$ is then on the same side as
monomer $i$ but slightly twisted by an angle $2\delta $ and so on. The
geometrical properties are given in Table I. For $\phi =\pi $, i.e., $\delta
=0$, we recover the result for the planar zig-zag structure.

The fiber is contained within a cylinder of radius $R$, given in Table I.
The monomers (''histones'') are located at the surface, with the linker
passing back and forth (approximately) through the middle axis of the
cylinder. The monomers $n,n\pm 2,n\pm 4...$ and the monomers $n\pm 1,n\pm
3,...$ form a double helix that winds around the cylinder. Within each of
the two spirals the monomers are not directly linked together, even though
monomer $i$ and $i+2$ can come quite close in space for large values of $%
\theta $. The pitch angle of the two spirals follows from the positions of
monomer 1 and 3; $P_{1}=\left( R,0,0\right) $ and $P_{3}\simeq \left(
R,-2R\delta ,2b\cos \left( \theta /2\right) \right) $, cf. Eq. \ref{sol}.
Thus $\gamma =-\pi +\Delta \gamma $ with $\Delta \gamma \simeq \delta \tan
\left( \theta /2\right) /2$.

\subsection{Structure diagram and excluded volume restriction}

We now consider the full range of states in the $\left( \theta ,\phi \right) 
$-space as shown in Fig. 4. Both angles $\theta $ and $\phi $ can each vary
over the range $0$ to $\pi $. At the edges of the diagram where one of the
angles assumes an extremal value, the configurations are always planar. On
the line $\phi =0$ we find circles and star-type polygons (that are closed
for specific values of $\theta $). The planar zig-zag-structures are located
on the line $\phi =\pi $; for $\theta =0$ we find a straight configuration
and for $\phi =\pi $ a ''dimer'' structure. If we move from the line $\phi
=0 $ towards larger values of $\phi $ the circles and star-like polygons
stretch out into the direction perpendicular to their plane, forming a
solenoid and a fiber with crossed linkers, respectively. On the other hand,
if we start at the top of the diagram ($\phi =\pi $) and decrease the value
of $\phi $ the planar zig-zag structure extends into the third dimension by
becoming twisted. If we start with a structure with entry-exit angle $\theta
=0$ and increase the value of this angle, then the structure folds either
into a solenoid with large pitch angle for small $\phi $-value or into a
twisted zig-zag for large values of $\phi $. Finally, starting out at the
dimer configuration, $\theta =\pi $ leads to an unfolding of the structure
into a fiber with crossed linkers (small $\phi $-values) or a twisted
zig-zag (large $\phi $-values).

If we take into account the excluded volume of the core particles, then
certain areas in our phase diagram are forbidden -- reminiscent of the
familiar ''Ramachandran plots'' used in the study of protein folding
(Stryer, 1995). For simplicity we assume in the following that the core
particles are spherical with a radius $a$ and that their centers are located
at the joints of two linkers, cf. Fig. 3. There are two different types of
interactions. One is between monomers at position $i$ and $i\pm 2$ (short
range interaction), and leads to the requirement that the entry-angle must
be sufficiently small: 
\begin{equation}
\theta <2\arccos \left( a/b\right) \simeq \pi -\frac{2a}{b}\text{,\thinspace
\thinspace \thinspace \thinspace \thinspace \thinspace \thinspace \thinspace 
}a\ll b  \label{ev1}
\end{equation}
This condition excludes a vertical strip at the right side of the diagram,
as indicated in Fig. 4 by a dashed line.

There is {\it also} a long-range excluded volume interaction that comes into
play when the angle $\phi $ is too small. This is apparent for the case $%
\phi =0$ where we find planar structures that run into themselves. Starting
with a circular structure we have to increase $\phi $ above some critical
value so that the pitch angle of the resulting solenoid is large enough so
that neighboring {\it loops} do not interact. This leads to the requirement $%
d>2a$ with $d$ given by Eq. \ref{d} (using $\phi \ll \theta $), i.e., 
\begin{equation}
\phi >\frac{1}{\pi }\frac{a}{b}\theta ^{2}  \label{ev2}
\end{equation}
For the large $\theta $-case (fibers with crossed linkers) we find from Eq. 
\ref{d2} the condition 
\begin{equation}
\phi >\frac{8}{\pi }\frac{a}{b}  \label{ev3}
\end{equation}
The two conditions, Eqs. \ref{ev2} and \ref{ev3}, shown schematically as a
dotted curve in Fig. 4, lead to a forbidden strip in the structure diagram
for small values of $\phi $.

Figure 4 does not show the interesting ''fine-structure'' of the boundary of
the forbidden strip that is due to commensurate-incommensurate effects. We
already noted that there are special $\theta $-values for which the
projection of the linkers forms a regular polygonal star ($\theta _{n}=\pi
\left( n-1\right) /n$) or a regular polygon ($\theta _{n}^{\prime }=2\pi /n$%
) (for small values of $\phi $). In these cases the nucleosomes $i$ and $i+n$
''sit'' on top of each other. On the other hand, for other values of $\theta 
$, monomers of neighboring loops will be displaced with respect to each
other. In this case monomers of one loop might be able to fill in gaps of
neighboring loops so that the minimal allowed value of $\phi $ is smaller
than estimated above. We have not explored the interesting mathematical
problem of the exact boundary line since this is likely to be sensitive to
the exact nucleosome shape. The dotted line in Fig. 4 only represents the
upper envelope of the actual curve.

Our discussion of the two-angle model was based of the assumption of a
perfectly homogeneous fiber where $b$, $\theta $ and $\phi $ are constant
throughout the fiber. The effect of randomness in these values on the fiber
geometry is discussed in Appendix B.

\section{Chromatin and the two-angle model: optimization of design?}

Where in the structure diagram is actual chromatin located? The classical
solenoid model of Finch and Klug (Finch and Klug, 1976) is found in the
small $\theta $, small $\phi $ section of the diagram (although in their
case the linker is bent). Various structures were displayed by Woodcock et
al. in their Fig. 2 (Woodcock et al., 1993), namely fibers with $\theta =150{%
{}^{\circ }}$ and many different values of $\phi $, corresponding to a
vertical trajectory on the right-hand side of Fig. 4. Three different
configurations with a fixed value of $\phi $ and different values of $\theta 
$ are displayed in Fig. 3(c) in another paper by these authors (Bednar et
al., 1998).

Our structure diagram accommodates all of these structures and, by itself,
does not favor one over another. However, our diagram plus the formulae
given above are useful if we invoke the following two criteria to optimize
the structure of the 30-nm fiber: 
\begin{eqnarray*}
&&({\it i})\,\,{\it maximum\ compaction} \\
&&({\it ii})\,\,{\it maximum\ accessibility}
\end{eqnarray*}
The first criterion is obvious: inactive chromatin should be packed as dense
as possible because of the very large ratio of DNA length to nucleus size.
By the second criterion we mean that a {\it local accessibility }mechanism
is required for gene transcription.

In order to attain maximum compaction we need structures that lead to high
bulk densities $\rho $ (we assume that the 30-nm fibers are packed in
parallel forming a hexagonal lattice). A comparison of the 3D densities of
the three different structures given in Table I shows that fibers with
internal linkers have highest densities $\rho $, namely 
\begin{equation}
\rho \simeq \frac{16}{\pi \phi \left( \pi -\theta \right) b^{3}}  \label{rho}
\end{equation}
In particular, the highest density is achieved for the largest possible
value of $\theta $ and the smallest possible value of $\phi $ that is still
in accordance with the excluded volume condition. This set of angles is
located at the point where the dotted curve and the dashed line in Fig. 4
cross each other. Apparently this {\it also} represents the only region in
the phase diagram where excluded volume effects are operative on a
short-range and a long-range scale at the {\it same} time, i.e., nucleosome $%
i$ is in close contact with nucleosome $i-2$ and $i+2$ {\it as well as} with
nuclesomes father apart along the contour length of the necklace. This
unique set of angles is given by $\theta _{\max }\approx 2\arccos \left(
a/b\right) $, cf. Eq. \ref{ev1}, and $\phi _{\min }\approx \left( 8/\pi
\right) \left( a/b\right) $, cf. Eq. \ref{ev3}.

In order to achieve maximum accessibility we look for structures that, for a
given entry-exit angle $\pi -\theta $ of a highly compacted structure,
achieve the maximum reduction in nucleosome line density $\lambda $ for a
given small change $\Delta \theta $ of the angle $\theta $. In other words,
we look for a maximum of $d\lambda /d\theta \,$which we call the
''accessibility''. {\it Interestingly, the accessibility is maximized at the
same unique pair of angles} $\left( \theta _{\max }\text{,}\phi _{\min
}\right) $. This can be seen from its angle dependence for fibers with
crossed linkers 
\begin{equation}
\frac{d\lambda }{d\theta }\simeq \frac{4}{\phi b\left( \pi -\theta \right)
^{2}}  \label{acc}
\end{equation}
We note that this change in $\lambda $ with $\theta $ is achieved by
changing the number of monomers per vertical repeat length $d$. The length $%
d $ itself is only weakly dependent on $n$ according to Eq. \ref{d2}.

Before we compare our theoretical formulas with experimental results we
mention that for fibers with crossed linkers there might be another excluded
volume interaction, namely between linkers. For these structures the linker
connecting the monomers $i$ and $i+1$ comes closest to the linkers between
monomer $i+2$ and $i+3$ and the one between $i-1$ and $i-2$, as can be seen
for the $n=5$ case, cf. ''5'' and ''10'' in Fig. 4. The linkers cross close
to the middle of the fiber where the distance between their axes is given by 
$2\left( \phi b/2\right) \cot \left( \theta /2\right) \simeq \phi b\left(
\pi -\theta \right) /2$, cf. Eq. \ref{d2}. This distance is minimized at $%
\left( \theta _{\max }\text{,}\phi _{\min }\right) $ and has to be larger
than the thickness $t$ of the fiber: 
\begin{equation}
t<\frac{8}{\pi }\frac{a^{2}}{b}  \label{link}
\end{equation}

We compare now the above given formulas with experimental results. For
chicken erythrocyte chromatin one has roughly $b\approx 20nm$
(center-to-center distance of nucleosomes, van Holde and Zlatanova, 1996).
Together with $a\approx 5nm$ this leads to $\theta _{\max }\approx 151{%
{}^{\circ }}$, $\phi _{\min }\approx $ $\allowbreak 36{{}^{\circ }}$ and $%
\lambda \approx 6.9$ nucleosomes per $11nm$ (cf. Eqs. \ref{ev1}, \ref{ev3}
and \ref{rho}). Furthermore, the condition on the linker thickness is given
by $t<3.2nm$ and is fulfilled since $t=2nm$ for DNA. The theoretically
derived values can now be compared with the ones reported by Bednar et al.
for chicken erythrocyte chromatin fibers (Bednar et al., 1998). From their
table 1we find that for an ionic strength of $80mM$ (which is close to the
physiological value) $\theta \approx 145{{}^{\circ }}$ and $\lambda =5.9$
nucleosomes per $11nm$. Furthermore, electron cryotomography-constructed
stereo pair images of an oligonucleosome (cf. Fig. 3(b) in Bednar et al.,
1998) indicate that the chromatin fiber might indeed have the structure of a
fiber with crossed linkers, with $n\approx 5$; this would correspond to $%
\theta =\pi \left( n-1\right) /n\approx 144{{}^{\circ }}$.

Information concerning the preferred value for $\phi $ may be obtainable
from the measured statistical distribution of the nucleosome repeat lengths.
This distribution shows statistically preferred linker lengths equal to $%
10k+1$bp's with $k$ a positive integer (Widom, 1992), which, in turn,
indicates that the rotation angle $\phi $ corresponds to a change in helical
pitch associated with $1$bp, i.e. $360^{\circ }/10=36^{\circ }$. This value
coincides with $\phi _{\min }$, the value that we estimated for maximum
compaction.\footnote{%
The statistical uncertainty around the expectation values for the nucleosome
repeat length is sufficiently large to make our estimate for $\phi $ less
reliable.}

The second feature, the local accessibility, can be monitored in vitro by
changing the salt concentration. Bednar et al. report, for example, that $%
\theta $ decreases with decreasing ionic strength, namely $\theta \approx 145%
{{}^{\circ }}$ at $80mM$, $\theta \approx 135{{}^{\circ }}$ at $15mM$ and $%
\theta \approx 95{{}^{\circ }}$ at $5mM$ (Bednar et al., 1998). In the
biochemical context the change of $\theta $ is accomplished by other
mechanisms, especially by the depletion of linker histones and the
acetylation of core histone tails (cf., for instance, van Holde and
Zlatanova, 1996), both of which are operative in transcriptionally active
regions of chromatin. These mechanisms lead effectively to a decrease of $%
\theta $.

As pointed out below Eq. \ref{acc}, the decrease of $\theta $ is accompanied
by a decrease of the line-density $\lambda =n/d$ of nucleosomes at an
essentially fixed value of $d$. In other words, the number of vertices of
the projected polygon decreases significantly with decreasing $\theta $
because $\theta _{n}=\pi \left( 1-1/n\right) $. In that respect the effect
of reducing $\theta $ below the optimal packing value might be best viewed
as an ''untwisting'' of the 30-nm fiber. Using the experimentally determined
values of $\theta $ we find from Table I that the density (the number of
nucleosomes per $11nm$) is given by $\lambda \approx 6.8$ for $\theta
\approx 145{{}^{\circ }}$, $\lambda \approx 4.5$ for $\theta \approx 135{%
{}^{\circ }}$ and $\lambda \approx 2.3$ for $\theta \approx 95{{}^{\circ }}$%
, slightly higher than the experimental values $\lambda \approx 6.0$, $%
\lambda \approx 3.2$ and $\lambda \approx 1.5$ (Bednar et al., 1998).
Furthermore, the number of polygonal vertices $n=\pi /\left( \pi -\theta
\right) $ decreases as follows: $n\approx 5.1$ for $\theta \approx 145{%
{}^{\circ }}$, $n\approx 4.0$ for $\theta \approx 135{{}^{\circ }}$ and $%
n\approx 2.1$ for $\theta \approx 95{{}^{\circ }}$, consistent with the
stereo pair images by Bednar et al., suggesting $n\approx 5$ at an ionic
strength of $80mM$ and $n\approx 3$ at $5mM$ (cf. Figs. 3(a) and (b) in
Bednar et al., 1998).

We close this section with a cautionary remark. The 3D density and the line
density of the fiber can not only be changed by changing $\theta $ or $\phi $
but also by changing the linker length (in multiples of $10$bp's). A
variation in $b$ changes the location of the point $\left( \theta _{\max
},\phi _{\min }\right) $ in the diagram of geometrical states, and thus the
values of the maximum 3D and line densities that can be achieved, namely 
\begin{equation}
\rho _{\max }\simeq \frac{16}{\pi \phi _{\min }\left( \pi -\theta _{\max
}\right) b^{3}}\simeq \frac{1}{a^{2}b}  \label{rhomax}
\end{equation}
and 
\begin{equation}
\lambda _{\max }\simeq \frac{4}{b\phi _{\min }\left( \pi -\theta _{\max
}\right) }\simeq \frac{\pi }{4}\frac{b}{a^{2}}  \label{lamdamax}
\end{equation}
This shows that fibers with smaller values of $b$ can achieve higher 3D
densities but have a smaller maximal line density (and accessibility $%
d\lambda /d\theta \propto b^{2}$). From this one might infer that active
cells should have larger nucleosome repeat lengths in order to maximize the
accessibility to their genetic material. An overview on nucleosome repeat
lengths in different organisms and tissues is given in table 7-1 of van
Holde's book (van Holde, 1989). The data shown there do not follow this
rule, unfortunately. In fact, very active cells like yeast cells and
neuronal cells have in general short nucleosome repeat lengths while
inactive ones like sperm cells have large ones. This shows that the
optimization principle of high density has to be used with caution.

\section{Stretching and compression of two-angle fibers}

\subsection{Introduction}

The $\left( \theta ,\phi ,b\right) $ model developed so far is purely
geometrical. Could it be useful as well for predicting {\it physical
properties} of the 30-nm fiber? The response of the 30-nm fiber to {\it %
elastic stress} will be the focus of this section. The elastic stress can
either be of external or of internal origin. External stresses are exerted
on the chromatin during the cell cycle when the mitotic spindle separates
chromosome pairs. The 30-nm fiber should be both highly flexible and
extensible to survive these stresses. The in vitro experiments by Cui and
Bustamante demonstrated that the 30-nm fiber is indeed very ''soft'' (Cui
and Bustamante, 2000).

The 30-nm fiber is also exposed to {\it internal stresses}. Attractive or
repulsive forces between the nucleosomes will deform the linkers connecting
the nucleosomes. For instance, electrostatic interactions, either repulsive
(due to the net charge of the nucleosome core particles) or attractive
(bridging via the lysine-rich core histone tails; Luger et al., 1997) could
lead to considerable structural adjustments of the $\left( \theta ,\phi
,b\right) $ model.

In this section we will derive an analytical description of the force
extension curve of the $\left( \theta ,\phi ,b\right) $ model in order to
predict the elastic properties of the different structures obtained in the
previous section. Using the particular values of $\theta $ and $\phi $ that
are observed experimentally (Bednar et al., 1998; Widom, 1992), we can
reproduce rather well the measured force-extension curve of Cui and
Bustamente and the numerical results of Katritch et al. that were based on a
variant of the $\left( \theta ,\phi ,b\right) $ model (see below).

Before considering the elastic properties of the $\left( \theta ,\phi
,b\right) $ model, it is helpful to briefly recall some results concerning
the large-scale elasticity of the DNA itself (Cluzel et al., 1996; Marko,
1998). The measured force-extension curve of naked DNA breaks up into two
highly distinct regimes: the ''entropic'' and ''enthalpic'' elastic regimes.
For very low tension $f$ ($\lesssim 1pN$), the restoring force is provided
by ''entropic elasticity'' (de Gennes, 1979). In the absence of any force
applied to its ends, the DNA's rms end-to-end distance (chain length, $L$)
is small compared to its contour length ($L_{0}$) and the chain enjoys a
large degree of conformational disorder. Stretching DNA reduces its entropy
and increases the free energy. The corresponding force $f$ increases
linearly with the extension $L$: 
\begin{equation}
f\simeq \frac{3k_{B}T}{A_{\text{DNA}}}\frac{L}{L_{0}},\,\,\,\,\,\,L\ll L_{0}
\label{dna1}
\end{equation}
The length $A_{\text{DNA}}$ is known as the ''thermal persistence length''
of DNA and is of the order $50nm$ (Hagerman, 1988).

For higher forces ($f\gtrsim 10pN$), the end-to-end distance $L$ is close to 
$L_{0}$ and the elastic restoring force is due to distortion of the internal
structure of DNA. In this regime, the force extension curve can be
approximated by 
\begin{equation}
f\simeq k_{B}T\gamma _{\text{DNA}}\frac{L-L_{0}}{L_{0}},\,\,\,\,\,\,L\gtrsim
L_{0}  \label{dna2}
\end{equation}
We will call $\gamma =\left( \partial f/\partial L\right) L_{0}/k_{B}T$ the
''stretching modulus''. $\gamma _{\text{DNA}}$ is about $300nm^{-1}$ (Cluzel
et al., 1996; Smith et al., 1996), i.e., almost four orders of magnitude
larger than the corresponding value $3/A_{\text{DNA}}$ obtained from Eq. \ref
{dna1}.

\subsection{Bending and twisting of linker}

To calculate the stretch modulus of the $\left( \theta ,\phi ,b\right) $
model, each linker is modeled as a wormlike chain (WLC) of fixed length $b$
(see Schlick, 1995 for a review of the WLC). We denote the geometrical
configuration of the $k$th linker ($k=1,2,3,...$) by ${\bf r}_{k}\left(
s\right) $ with $s$ being the arclength, $0\leq s\leq b$. The elastic energy
of the wormlike linker is given by the sum of the bending and the torsional
energies: 
\begin{equation}
E_{k}=\frac{1}{2}\int_{0}^{b}ds\left\{ \kappa \left( \frac{1}{R_{k}\left(
s\right) }\right) ^{2}+C\left( \frac{d\eta _{k}\left( s\right) }{ds}\right)
^{2}\right\}  \label{en}
\end{equation}
Here $\kappa $ is the bending stiffness which is related to the persistence
length $A_{\text{DNA}}$ of (linker) DNA by $\kappa =k_{B}TA_{\text{DNA}}$.
Furthermore, $1/R_{k}\left( s\right) =\left| d^{2}{\bf r}_{k}\left( s\right)
/ds^{2}\right| $ denotes the curvature of the $k$th linker at the point $s$
along its contour. The torsional angle of the linker is $\eta _{k}\left(
s\right) $ and the torsional stiffness is $C$. The positions ${\bf r}%
_{k}\left( 0\right) $ and ${\bf r}_{k}\left( b\right) $ of the two termini
of the $k$th linker coincide with the termini of the neighboring linkers,
i.e., ${\bf r}_{k-1}\left( b\right) ={\bf r}_{k}\left( 0\right) $ and ${\bf r%
}_{k}\left( b\right) ={\bf r}_{k+1}\left( 0\right) $. Furthermore, we assume
that the entry-exit angles have the fixed value $\pi -\theta $ independent
of the bending and twisting of the linkers. This means that the unit
tangents fulfill the condition $\cos \left( \theta \right) ={\bf t}%
_{k}\left( b\right) \cdot {\bf t}_{k+1}\left( 0\right) $, with ${\bf t}%
_{k}\left( s\right) =d{\bf r}_{k}\left( s\right) /ds$.

({\it i}) {\it Enthalpic elasticity}: We study first the stretching of the 
{\it planar zig-zag pattern} ($\phi =\pi $, $\theta $ arbitrary). The
undeformed zig-zag fiber is depicted in Fig. 5(a). In order to give a more
accurate description of the mechanical properties of the fiber, we assume
that the nucleosome core particles are not located at the sites where two
linkers come together but rather slightly displaced, forming a stem
configuration as shown in Fig. 5(a). This is the geometry obtained from the
electron cryomicrographs of Bednar et al. (1998) and it is the same geometry
that was adopted in the computer simulations of fiber stretching (Katritch
et al., 2000). In the following we denote the actual linker length by $\bar{b%
}$\ in order to distinguish it from $b$, the distance between neighboring
nucleosomes. For symmetry reasons, there is no torque on the structure, so
that $d\eta _{k}/ds\equiv 0$. The stretching of the fiber is achieved by a
bending of the linkers with the entry-exit-angle $\theta $ $\,$remaining
constant, cf. Fig. 5(b). This leads to a deformation where the tangent
vectors ${\bf t}_{k}\left( 0\right) $ and ${\bf t}_{k}\left( \bar{b}\right) $
remain parallel but undergo {\it lateral displacement}. We assume a small
deformation of the linker with displacement $u\left( s\right) $ from the
straight configuration small compared to $\bar{b}$. (Since $u\left( s\right) 
$ is the same for all the linkers, we drop the index $k$ from here on.)

From the minimization of $E_{k}$, Eq. \ref{en}, we obtain the Euler-Lagrange
equation $d^{4}u/ds^{4}=0$. The boundary conditions that must be obeyed by
the solutions are $u\left( 0\right) =u^{\prime }\left( 0\right) =u^{\prime
}\left( \bar{b}\right) =0$ and $u\left( \bar{b}\right) =d$ where $d$
describes the displacement of the bead vertical to the original straight
linker (we assume $d\ll \bar{b}$ here and neglect terms of the order $\left(
d/\bar{b}\right) ^{2}$). It follows that the deformation profile is given by 
$u\left( s\right) =-2ds^{3}/\bar{b}^{3}+3ds^{2}/\bar{b}^{2}$. The associated
bending energy is $E=6\kappa d^{2}/\bar{b}^{3}$ (per linker). The
deformation translates into an effective change in the deflection angle from 
$\theta $ to $\theta -\Delta \theta $ where $\Delta \theta /2=d/\bar{b}$ --
see Fig. 5(b). The energy of a fiber with $N$ linkers as a function of $%
\Delta \theta $ is thus given by $E=\left( 3/2\right) \left( \kappa /\bar{b}%
\right) \left( \Delta \theta \right) ^{2}N$.

The change in $\theta $ produces a change in the overall length of the
fiber. We find from Eq. \ref{L}:

\begin{equation}
L=\bar{b}N\cos \left( \frac{\theta -\Delta \theta }{2}\right) \simeq L_{0}+%
\bar{b}N\sin \left( \theta /2\right) \Delta \theta /2,\,\,\,\,\,\Delta
\theta \ll 1\,\left( \leftrightarrow d\ll \bar{b}\right)  \label{L5}
\end{equation}
where $L_{0}$ is the contour length of the unperturbed fiber, Eq. \ref{L}.
The energy can be rewritten in terms of the extension $\Delta L=L-L_{0}$.
The restoring force follows then from $f=dE/dL$: 
\begin{equation}
f\simeq \frac{12}{N\sin ^{2}\left( \theta /2\right) }\frac{\kappa }{\bar{b}%
^{3}}\Delta L  \label{force2}
\end{equation}
The associated stretching modulus (defined as in Eq. \ref{dna2}) follows
from Eqs. \ref{L} and \ref{force2}: 
\begin{equation}
\gamma _{fiber}\left( \theta \right) \simeq \frac{12A_{\text{DNA}}}{\bar{b}%
^{2}}\frac{\cos \left( \theta /2\right) }{\sin ^{2}\left( \theta /2\right) }
\label{gamma}
\end{equation}

We next consider the deformation of {\it fibers with crossed linkers} ($\phi
\ll 1$, $\theta $ large){\it .} When such a fiber is stretched, linkers will
be twisted as well as bent. Interestingly, a fiber with a high bending
stiffness ($\kappa \rightarrow \infty $) can still be stretched just by
twisting of the linkers. When one applies a tension to such a fiber each
linker is twisted and the rotational angle changes by $\Delta \phi $ from
one linker to the next, i.e., $d\eta /ds=\Delta \phi /\bar{b}$. The twist is
distributed homogeneously along the linker since $d^{2}\eta /ds^{2}=0$ which
follows from minimization of $E_{k}$ in Eq. \ref{en}. The energy per linker
is given by $E=\left( C/2\right) \left( \Delta \phi ^{2}/\bar{b}\right) $.
The twist of the linkers changes the length of the fiber, and using $L$ from
Table I it is straightforward to calculate the force as a function of the
relative extension: 
\begin{equation}
f\simeq \frac{4C}{\cot ^{2}\left( \theta /2\right) \bar{b}^{3}N}\Delta L
\label{force1}
\end{equation}
In the opposite limit of linkers with extremely high torsional stiffness ($%
C\rightarrow \infty $), the force-extension curve can be mapped onto the
planar zig-zag case, described above, by replacing $\pi -\theta $ by $\tilde{%
\phi}\simeq \phi \cot \left( \theta /2\right) $, the ''effective'' angle
between two consecutive linkers as seen from the ''side'' of the fiber.
(This follows from $\tilde{\phi}\simeq 2\Delta l/\bar{b}$ where $\Delta l$
is the difference in the longitudinal position of bead $i$ and $i+1$; $%
\Delta l=\left( \phi \bar{b}/2\right) \cot \left( \theta /2\right) $) Using
Eq. \ref{force2} we find 
\begin{equation}
f\simeq \frac{12\kappa }{\bar{b}^{3}N\cos ^{2}\left( \tilde{\phi}/2\right) }%
\Delta L  \label{force5}
\end{equation}

We can now define as before the two stretching moduli $\gamma _{twist}$ and $%
\gamma _{bend}$ using Eqs. \ref{force1} and \ref{force5} together with Eq. 
\ref{d2}. If we allow {\it both} twist and bend then the two ''spring
constants'' act ''in series'': $\gamma _{fiber}^{-1}=\gamma
_{twist}^{-1}+\gamma _{bend}^{-1}$. We obtain for the stretching constant of
the fiber ($\phi \ll 1$, $\theta $ large) 
\begin{equation}
\gamma _{fiber}\left( \theta ,\phi \right) =\gamma _{bend}\left( 1+\frac{%
\gamma _{bend}}{\gamma _{twist}}\right) ^{-1}=\frac{6A_{\text{DNA}}}{\bar{b}%
^{2}}\frac{\tilde{\phi}}{\cos ^{2}\left( \frac{\tilde{\phi}}{2}\right) }%
\left[ 1+\frac{3\cot ^{2}\left( \frac{\theta }{2}\right) }{\cos ^{2}\left( 
\frac{\tilde{\phi}}{2}\right) }\frac{\kappa }{C}\right] ^{-1}  \label{gamma2}
\end{equation}
For large $n$ (with $n$ defined as $n=\pi /\left( \pi -\theta \right) $) the
twisting contribution can be neglected since then $\gamma _{bend}/\gamma
_{twist}\approx \left( 3/4\right) \left( \pi /n\right) ^{2}\left( \kappa
/C\right) $. For DNA $C\simeq k_{B}T\times 750$A$\gtrsim \kappa $ ( Klenin
et al., 1989; Crothers et al., 1992) so that $\gamma _{bend}\ll \gamma
_{twist}$ for $n\gtrsim 5$. In this case one has $\gamma _{fiber}\simeq
\gamma _{bend}$ and Eq. \ref{force5} applies.

({\it ii}) {\it Entropic elasticity}: Just as for naked DNA, the entropic
contribution to the elasticity dominates for weak forces ($L\ll L_{0}$). The
restoring force is again of the form 
\begin{equation}
f=\frac{3k_{B}T}{A_{fiber}}\frac{L}{L_{0}}  \label{entropic}
\end{equation}
with $A_{fiber}$ the persistence length of the fiber. This persistence
length is calculated in Appendix C. For the case of the crossed-linker fiber
we find 
\begin{equation}
A_{fiber}\approx A_{\text{DNA}}\frac{\phi }{2}\cot \left( \theta /2\right)
\label{afiber}
\end{equation}
For values of $\theta $ and $\phi $ appropriate for the crossed-linker
structure, $A_{fiber}$ {\it is somewhat less than }$A_{\text{DNA}}$ (of the
order of $50nm$). This surprising conclusion is related to the fact that a
large amount of DNA material is stored in the fiber per unit length. The
30-nm fiber is thus indeed highly flexible.

Our calculation of the stretching properties of the two-angle model predicts
an important difference between the stretching behavior of DNA and that of
the 30-nm fiber. The enthalpic stretching modulus of a chromatin fiber is of
the order $A_{\text{DNA}}/\bar{b}^{2}$ (see Eqs. \ref{gamma} and \ref{gamma2}%
) which is about $0.3nm^{-1}$ for linkers with $40$bp's ($\bar{b}=40\times
0.34nm\simeq 14nm$). This is only{\it \ }an order of magnitude larger than
the entropic stretching modulus $1/A_{fiber}\simeq 0.03nm^{-1}$. In other
words, because of the low value of $\gamma _{fiber}$ and because of $%
A_{fiber}\approx A_{\text{DNA}}$ there is {\it no longer a very clear
distinction between entropic and enthalpic behavior} as it is observed for
naked DNA. In conclusion, the 30-nm fiber shows {\it soft elasticity} under
stretching due to bending and twisting of the linkers.

\subsection{Internucleosomal attraction}

The effect of attractive interaction between nucleosomes is to cause a {\it %
compression} of the 30-nm fiber. Phase behavior studies of linker-free
nucleosome solutions, i.e.,{\it \ }solutions of disconnected nuclesomes
(Livolant and Leforestier, 2000, cf. also Fraden and Kamien, 2000) indicate
that nucleosome core particles spontaneously form fiber-like {\it columnar
structures}, presumably due to attractive nucleosome-nucleosome interaction.
Attractive nucleosome interaction could be mediated for instance by the
lysine-rich core histone tails (Luger et al., 1997), as mentioned above.

It is important to distinguish these condensed fibers from the swollen
solenoid-, zig-zag- and crossed-linker structures predicted by the $\left(
\theta ,\phi ,b\right) $-model. The dominant energy of the condensed
structures is the nucleosome attractive interaction, while the ''swollen''
structures are dominated by linker elasticity. In this section we will
discuss the competition between swollen and condensed phases for a simple
case.

For simplicity, we model the fiber as a planar zig-zag structure with
elastic linkers and assume in addition a short-range interaction between
nucleosomes. This interaction, denoted by $U_{inter}$, is assumed to be a
short range attraction, of strength $-U_{min}$, that acts only when the
nucleosomes are in close contact, i.e., at a distance $x\approx 2a$ of the
order of the hardcore diameter. For a given nucleosome, say the $i$th, the
closest nucleosomes in space are number $i+2$ and $i-2$ as discussed in
Section 2. We will disregard the interaction between other pairs. The
elastic interaction $U_{el}$ follows directly from Eq. \ref{force2} with $%
N=2 $: 
\begin{equation}
U_{bend}\left( x\right) =\frac{3}{\sin ^{2}\left( \theta /2\right) }\frac{%
\kappa }{\bar{b}^{3}}\left( x-x_{0}\right) ^{2}=\frac{K}{2}\left(
x-x_{0}\right) ^{2}  \label{ubend}
\end{equation}
where $x_{0}=2\bar{b}\cos \left( \theta /2\right) $ denotes the distance
between nucleosome $i$ and $i+2$ for straight linkers (cf. Eq. \ref{L}). The
total internucleosomal $U\left( x\right) $ equals $U_{inter}\left( x\right)
+U_{bend}\left( x\right) $.

Fig. 6(a) shows $U\left( x\right) $ for different values of $\theta $. We
assume for simplicity that the interaction energy $U_{inter}$ remains
unchanged. Curve ''1'' in Fig. 6(a)) shows $U\left( x\right) $ for a small
value of $\theta $ where the global minimum of $U\left( x\right) $ is
located at $x=x_{0}$ denoted by ''S'' (swollen state). Curve ''2''
corresponds to an intermediate value of $\theta $ at which the minima at
''S'' and ''C'' have the same value. For this value of $\theta $, $\theta
=\theta _{c}$, the energy minimum shifts from ''S'' to a new minimum,
representing the condensed state ''C''. The change in $\theta $ produced a 
{\it structural transition} from a swollen state to a condensed state.
Finally, curve ''3'' depicts $U\left( x\right) $ for a deflection angle $%
\theta >\theta _{c}$ with the minimum at ''C''. The critical angle for the
''S'' to ''C'' transition can be determined by comparing the bending energy
at close contact, $U_{bend}\left( 2a\right) $, and the strength $U_{min}$ of
the short range attraction. Equating both leads to the following condition
for $\theta _{c}$: 
\begin{equation}
\cos \left( \theta _{c}/2\right) -\sqrt{\frac{\bar{b}\left(
U_{min}/kT\right) }{6A_{\text{DNA}}}}\sin \left( \theta _{c}/2\right) =\frac{%
a}{\bar{b}}  \label{thetac}
\end{equation}
In the swollen state the elastic properties are those discussed in the
previous section. In the condensed state, the elastic properties are
determined by the detailed form of the nucleosome interaction potential.

If the condensed state has a lower free energy, i.e. if $\theta >\theta _{c}$%
, then an external stretching force $f$ can induce a transition from the
condensed to the swollen state. The transition point $f_{CS}$ follows from a
''common-tangent'' construction. The conditions are $U^{\prime }\left(
x_{1}\right) =U^{\prime }\left( x_{2}\right) =f_{CS}$ and $\left( U\left(
x_{2}\right) -U\left( x_{1}\right) \right) /\left( x_{2}-x_{1}\right)
=f_{CS} $ (cf. Fig. 6(a)). The first pair of conditions leads to $x_{1}=2a$, 
$x_{2}=x_{0}+f_{CS}/K$. The last condition leads to 
\begin{equation}
f_{CS}=\sqrt{2KU_{min}}-K\left( x_{0}-2a\right)  \label{fcs}
\end{equation}

The corresponding force-extension curve has a ''coexistence plateau'', cf.
Fig. 6(b). If the imposed end-to-end distance is smaller than $L_{0}$ (the
contour length of the condensed fiber) then the restoring force is entropic.
For $L_{0}<L<L_{1}$ the force rises sharply with increasing $L$. This ''hard
elasticity'' is governed by the nucleosomal interaction potential $U_{inter}$%
. Then at $L=L_{1}$ the coexistence plateau is reached. Between $L=L_{1}$
and $L=L_{2}$ parts of the fiber are in the ''S'' state and parts are in the
''C'' state. For larger extensions, $L>L_{2}$, the fiber shows soft
elasticity due to the bending (and twisting ) of the linkers as discussed in
the previous section.

\subsection{Stretching chromatin}

We now compare the results of the previous two sections with the force
extension curves found in recent experiments (Cui and Bustamante, 2000).

({\it i}) {\it Low ionic strength}: We start with the force-extension
profile measured at low ionic strength (5mM), cf. Fig. 2 in Cui and
Bustamante. As discussed above, at low ionic strength the chromatin fiber
constitutes a swollen fiber with crossed linkers. The nucleosomes are far
apart and we assume that there is no direct interaction between nucleosomes.
The resulting force-extension profile is expected to show a crossover
between an entropic elasticity (cf. Eq. \ref{entropic}) and a soft enthalpic
elasticity with a stretching modulus given by Eq. \ref{gamma2}: 
\begin{equation}
f\simeq \left\{ 
\begin{array}{ll}
\frac{6k_{B}T}{l_{P}\bar{b}\tilde{\phi}N}L & \mbox{for}\;L\ll L_{0} \\ 
\frac{2k_{B}T\gamma _{fiber}}{\tilde{\phi}\bar{b}N}\left( L-L_{0}\right) +%
\frac{3k_{B}T}{l_{P}} & \mbox{for}\;L\gg L_{0}
\end{array}
\right.  \label{fregimes}
\end{equation}
with $L_{0}\simeq \left( \tilde{\phi}/2\right) \bar{b}N$, cf. Table 1.

Cui and Bustamante estimate the number of nucleosomes in their fibers to be $%
N\approx 280$. From the formula for $L$ in Table I, we would estimate the
length of the fiber to be $L_{0}\approx 1.0\mu m$ using the values $\theta
=95{{}^{\circ }}$ (Bednar et al., 1998), $\phi =36{{}^{\circ }}$ (Widom,
1992), and $\bar{b}=40bp=40\times 0.34nm=1.4\times 10^{-8}m$. The linker
length is estimated from the nucleosome repeat length of roughly $210$ bp's
(cf. Table 7-1F in van Holde, 1989) minus roughly $170$ bp's that are
associated with the core and linker histones (cf. page 268 in van Holde,
1989). Using the moduli for DNA (Hagerman, 1988), $\kappa =k_{B}T\times
50nm=2\times 10^{-16}pNm^{2}$, $C=k_{B}T\times 75nm=3\times 10^{-16}pNm^{2}$
and $l_{P}=30nm$ for the persistence length of the fiber (cf. Appendix C) we
find from Eq. \ref{fregimes} the following force-extension relation (force
in\thinspace $pN$, extension in $\mu m$):

\begin{equation}
f\simeq \left\{ 
\begin{array}{ll}
0.35\times L & \mbox{for}\;L\ll 1.1 \\ 
1.2\times \left( L-1.0\right) +0.4 & \mbox{for}\;L\gg 1.1
\end{array}
\right.  \label{forceexp}
\end{equation}
The agreement with the experimental curve at low ionic strength ($5mM$ NaCl)
is reasonable (cf. Fig. 2(a) and (b) in Cui and Bustamante, 2000). More
explicitly, for forces up to $5pN$ and extensions up to $\approx 2\mu m$
there are two distinctive regimes: For small extensions, $L\lesssim 1\mu m$,
the force increases only slightly with tension, namely roughly as $f\approx
0.5\times L$. Then for $L\gtrsim 1\mu m$ the measured force increases much
faster and shows the following linear dependence: $f\approx 7\times L$ (Fig.
2(a)) or $f\approx 5\times L$ (Fig. 2(b)). The different slopes in this
regime are a result of a slight hysteresis: the relaxation curve has a
smaller slope after the fiber has been stretched to an end-to-end distance $%
2.5\mu m$ (Fig. 2(b)) than for the case of a much smaller stretching cycle
(up to $1.8\mu m$, Fig. 2(a) in Cui and Bustamante, 2000); the hysteresis
disappears for a smaller rate of extension or contraction, and might be a
result of nucleosome-nucleosome interaction or of modifications of the fiber
close to the entry-exit point of the linkers at higher tension. We also
mention that for forces beyond $\approx 5pN$ the (relaxation) curve shows an
increasing slope, probably due to nonlinear effects not accounted for in the
current study\footnote{%
Our calculation is based on the assumption of small deformations. For the
zig-zag case this requires $d\ll \bar{b}$, i.e., $\Delta \theta \ll 1$, cf.
Fig. 5(b). Using Eqs. \ref{L5} and \ref{force2} this condition translates
into the requirement that the tension $f$ is smaller than $6\kappa /\left( 
\bar{b}^{2}\sin \left( \theta /2\right) \right) $. For fibers with internal
linkers the condition is $\Delta \tilde{\phi}\ll 1$, leading to $f\ll
6\kappa /\left( \bar{b}^{2}\cos ^{2}\left( \tilde{\phi}/2\right) \right) $.
Thus in both cases a good estimate for the range of forces where the linear
approximation holds is given by $f<6\kappa /\bar{b}^{2}$. For the chromatin
fiber under consideration we find $6\kappa /\bar{b}^{2}\simeq 6pN$.}.

The calculated forces are smaller than the measured ones (roughly by a
factor of $4$), for several reasons. First, the (mean) values of $\theta $, $%
\phi $ and $\bar{b}$ (and thus $N$) are only roughly known. Secondly, the
value of $\theta $ we used ($95{{}^{\circ }}$) is not large enough compared
to $\pi $ for the above given theoretical formulas to hold accurately.
However, as a check of our analytical approximations, we compared our
results with the computer simulations by Katritch et al. (Katritch et al.,
2000) where $\theta $, $\phi $ and $\bar{b}$ are variable. This comparison
is given in Appendix D, where we show that there is good agreement
indicating that our analytical approximations were in fact reasonable.

({\it ii}) {\it High ionic strength}: For 40mM NaCl or higher ionic strength
the chromatin fiber is much denser and nucleosomes approach each other
closely. Attractive short-range forces and the increase of $\theta $
associated with higher ionic strength should favor the condensed phase. A
plateau indeed appears at $5pN$ in the force-extension plot (cf. Fig. 4 in
Cui and Bustamante, 2000). From the extent of the plateau, $0.6\mu m$, its
height, $5pN$, and the number of nucleosomes in the stretched fiber, $%
\approx 280$, it was estimated that there is an attractive interaction
energy of roughly $3kT$ per nucleosome (Cui and Bustamante, 2000).

We now can use Eq. \ref{fcs} to estimate independently the strength of the
nucleosomal attraction from the value of the critical force alone. We find:

\begin{equation}
U_{min}=\frac{\left( f_{CS}+K\left( x_{0}-2a\right) \right) ^{2}}{2K}
\label{umin}
\end{equation}
If we neglect the second term in the bracket, we find $U_{min}\approx
f^{2}/\left( 2K\right) \approx 6kT$ (assuming $\theta =140{^{\circ }}$),
close to the value $3kT$ estimated directly from the force-extension diagram
(Cui and Bustamante, 2000) and also in accordance with the computer
simulation of Katritch et al. who obtained an internucleosomal short-range
attraction of order $2kT$ (Katritch et al., 2000).

Using $U_{min}=3kT$ we can estimate the critical value $\theta =\theta _{c}$
at which the condensed and the swollen chromatin fiber should coexist. We
find numerically from Eq. \ref{thetac} that $\theta _{c}\approx 100{^{\circ }%
}$ (using $a=5nm$ and $\bar{b}=14nm$). This value is lower than the one that
can be inferred from experiments. At $15mM$ NaCl ($\theta \approx 135^{\circ
}$) the fiber appears to be decondensed, as indicated by stretching
experiments and from electron cryomicrographs. This fact as well as the
appearance of a plateau in the force extension curve at $40mM$ salt (where $%
\theta \approx 140^{\circ }$) indicates that one should expect $135^{\circ
}\lesssim \theta _{c}\lesssim 140{{}^{\circ }}$ (cf. Bednar et al., 1998).
It should be recalled, however, that our model for the attractive
interaction is highly oversimplified.

\section{Conclusion}

The present analytical study of the $\left( \theta ,\phi ,b\right) $ model
first of all shows that this model can account for the measured
force-extension curve of the 30-nm fiber in the low-salt regime with, in
effect, {\it no fitting parameters} (since $\theta $, $\phi $, and $b$ can
be estimated experimentally and since the elastic moduli characterizing
naked DNA are known). Since the $\left( \theta ,\phi ,b\right) $ model also
accounts for the observed low-salt structure of the 30-nm fiber (''crossed
linkers''), there seems to be good evidence that this model is at least the
proper description in the {\it low-salt} regime.

We have been able to compute the structural and elastic properties over a
wide range of $\left( \theta ,\phi \right) $-values. We suggest that the
native chromatin fiber might be a particular realization of this rich array
of structures, namely the one that simultaneously maximizes compaction and
accessibility, consistent with the restriction of excluded volume between
nucleosomes.

Confirmation that a certain optimization principle is in fact operative for
biomolecules is usually a difficult issue. We already saw that, at best, the
principle is incomplete since the linker-length $b$ evidently is not
determined by the conditions of maximum compaction and accessibility. One
possibility may be to explore the fine-structure of the dotted curve in Fig.
4, the lower bound of $\phi $ as a function of $\theta $ . This is expected
to have an ''irregular'' shape due to commensurate-incommensurate effects
and it may be possible to associate a {\it discrete} geometrical structure
(e.g. a particular index $n$ for the polygonal star projection) with maximum
compaction and accessibility. Such a study would require, however, a better
description of the structure of individual nucleosomes and extensive
numerical work.

How confident can we be that the $\left( \theta ,\phi ,b\right) $ model is
appropriate as well in the biologically relevant regime of physiological
salt concentrations? We had to include a weak attractive nucleosome
interaction to explain the coexistence in the force-extension curve. If the
fitted value for the attractive potential ($U_{min}$) is used in Eq. \ref
{thetac} we obtain a reasonable estimate for the critical angle $\theta _{c}$
for the ''S'' to ''C'' transition (but with a significant error).

A completely different approach would be that the high-salt regime is
controlled not by a balance between soft elasticity and weak attraction but
completely by nucleosome-nucleosome attraction forces (plus short-range
repulsion). As shown by the work of Livolant and Leforestier (2000),
nucleosome attraction indeed can produce discoidal fiber structures (formed
by {\it linker-free} core particles) all by itself. If the interaction
energy is strong enough, then the linkers would be strongly bent in the
condensed state. The $\left( \theta ,\phi ,b\right) $ model would not be a
valid description anymore. The effect of tension could be to produce a
sequence of different condensed structures. Only at high tension when the
internucleosomal contacts are broken one recovers the soft-elasticity
regime, described well by the $\left( \theta ,\phi ,b\right) $ model. Which
of the two approaches is valid is an issue that must be determined
experimentally.

Interesting questions for ''chromatin physics'' in the future may focus on
dynamical issues. Suppose that $\theta $ is {\it locally }increased, e.g. by
acetylation of core histone tails, how long does it take for the
accessibility to increase sufficiently. How important is nucleosome mobility
(Schiessel et al., 2000) and nucleosome ''evaporation'' (Marko and Siggia,
1997) for the swelling dynamics of chromatin?

\acknowledgments
We wish to thank J. Widom for many valuable discussions and for a critical
reading of the manuscript. We would like to acknowledge useful conversations
with J.-L. Sikorav and F. Livolant. This work was supported by the National
Science Foundation under Grant DMR-9708646.

\newpage

\begin{center}
{\bf APPENDIX\ A:\ THE\ MASTER\ SOLENOID}
\end{center}

For any given set of angles $\left( \theta ,\phi \right) $ there is a
solenoid so that the successive monomers of the fiber structure lie
successively on this helical path. (There are actually many such solutions,
but we are interested in the one with the largest pitch angle $\gamma $.) We
parametrize the solenoid as follows 
\begin{equation}
{\bf r}\left( s\right) =\left( 
\begin{array}{l}
R\cos \left( \alpha s/R\right) \\ 
R\sin \left( \alpha s/R\right) \\ 
s
\end{array}
\right)  \label{sol}
\end{equation}
$R$ denotes the radius of the solenoid and $\alpha $ is related to the pitch 
$\gamma $ by 
\begin{equation}
\alpha =\cot \gamma  \label{alpha}
\end{equation}
(as follows from ${\bf \dot{r}}\left( 0\right) =\left( 0,\alpha ,1\right) $).

Assume now an infinite fiber of monomers with a given pair of angles $\left(
\theta ,\phi \right) $. The monomers are located at the positions ${\bf R}%
_{0},{\bf R}_{\pm 1},{\bf R}_{\pm 2},...$ The axis of the fiber coincides
with the $Z$-axis. Assume further that we choose the values $R$ and $\alpha $
so that the solenoid curve goes through all monomers. Put the monomer
labeled $i=0$ at $s=0$ so that ${\bf R}_{0}=\left( R,0,0\right) $; the
subsequent monomer, $i=1$, is at a position ${\bf R}_{1}$ given by Eq. \ref
{sol} with $s=s_{0}$. The next monomer is located at ${\bf R}_{2}={\bf r}%
\left( 2s_{0}\right) $. Finally, the position of monomer $i=-1$ is given by $%
{\bf R}_{-1}={\bf r}\left( -s_{0}\right) $.

Now let us calculate the bond vectors between these monomers. Monomer $i=1$
is connected to monomer $i=0$ via 
\[
{\bf r}_{0}={\bf R}_{1}-{\bf R}_{0}=\left( 
\begin{array}{l}
R\cos \left( \alpha s_{0}/R\right) -R \\ 
R\sin \left( \alpha s_{0}/R\right) \\ 
s_{0}
\end{array}
\right) 
\]
The separation vector between monomer $i=2$ and $i=1$ is given by 
\[
{\bf r}_{1}={\bf R}_{2}-{\bf R}_{1}=\left( 
\begin{array}{l}
R\left( \cos \left( 2\alpha s_{0}/R\right) -\cos \left( \alpha
s_{0}/R\right) \right) \\ 
R\left( \sin \left( 2\alpha s_{0}/R\right) -\sin \left( \alpha
s_{0}/R\right) \right) \\ 
s_{0}
\end{array}
\right) 
\]
and that between monomer $i=0$ and $i=-1$ by 
\[
{\bf r}_{2}={\bf R}_{0}-{\bf R}_{-1}=\left( 
\begin{array}{l}
R-R\cos \left( \alpha s_{0}/R\right) \\ 
R\sin \left( \alpha s_{0}/R\right) \\ 
s_{0}
\end{array}
\right) 
\]

$s_{0}$ follows from the condition of fixed linker length, i.e., $\left| 
{\bf r}_{0}\right| =b$. This leads to the relation 
\begin{equation}
b^{2}=2R^{2}\left( 1-\cos \left( \alpha s_{0}/R\right) \right) +s_{0}^{2}
\label{s0}
\end{equation}
We determine $\theta $ from $\cos \theta ={\bf r}_{0}\cdot {\bf r}%
_{2}/\left| r_{0}^{2}\right| $, which leads to 
\begin{equation}
\cos \theta =\frac{2R^{2}\cos \left( \alpha s_{0}/R\right) \left( 1-\cos
\left( \alpha s_{0}/R\right) \right) +s_{0}^{2}}{2R^{2}\left( 1-\cos \left(
\alpha s_{0}/R\right) \right) +s_{0}^{2}}  \label{cost}
\end{equation}
Finally, $\phi $ is the angle between normal vectors of the planes that are
defined by monomers 0 and 1, i.e. $\cos \phi ={\bf n}_{1}\cdot {\bf n}_{2}$.
We obtain ${\bf n}_{1}$ and ${\bf n}_{2}$ from ${\bf n}_{1}={\bf A}/\left| 
{\bf A}\right| $ and ${\bf n}_{2}={\bf B}/\left| {\bf B}\right| $ where $%
{\bf A}={\bf r}_{0}\times {\bf r}_{1}$ and ${\bf B}={\bf r}_{2}\times {\bf r}%
_{0}$. After some algebra we arrive at 
\begin{equation}
\cos \phi =\frac{s_{0}^{2}\cos \left( \alpha s_{0}/R\right) +R^{2}\sin
^{2}\left( \alpha s_{0}/R\right) }{s_{0}^{2}+R^{2}\sin ^{2}\left( \alpha
s_{0}/R\right) }  \label{cosf}
\end{equation}
Equations \ref{s0}, \ref{cost} and \ref{cosf} relate $\alpha $ (or $\gamma $%
), $R$ and $s_{0}$ of the spiral to $\phi $, $\theta $ and $b$.\bigskip

\begin{center}
{\bf APPENDIX\ B:\ RANDOMNESS\ IN\ THE\ }${\bf \phi }${\bf -DISTRIBUTION}
\end{center}

\smallskip Up to now we have assumed that the values of the angles $\theta $
and $\phi $ are constant throughout the fiber. The resulting ''ground
state'' configuration (unbent and untwisted linkers) is a fiber whose axis
is perfectly straight. The assumption that the linker entry-exit angle $%
\theta $ is constant is based on the fact that it is a local property of the
nucleosome core particle, and as long as the biochemical conditions are
homogeneous throughout the fiber this should be a reasonable assumption. It
is known, however, that the rotational positioning is not perfect, as can be
seen from the experimentally determined distribution of the linker length in
chromatin (Widom, 1992). Even though a preferred rotational setting can be
deduced, the width of the distribution of linker lengths will be reflected
in the width of the distribution of the angle $\phi $. If the rotational
setting of the nucleosomes were completely random, then the chromatin
configurations would correspond to particular configurations of the freely
rotating chain (if we neglect excluded volume effects) (Doi and Edwards,
1986). These configurations, in turn, are those of a Gaussian chain with a
persistence length $l_{P}=b\left( 1+\cos \theta \right) /\left( 1-\cos
\theta \right) $ (the Kuhn statistical length as defined in Doi and Edwards,
1986). Note that $l_{P}$ increases when $\theta $ decreases, a mechanism
similar to the accordion-like unfolding of the zig-zag structure or the
untwisting of the fiber with crossed linkers discussed above. In the
following we will assume small variations of the rotational setting around
some mean value $\phi $. We consider the three cases: the solenoid, the
fiber with crossed linkers, and (twisted) zig-zag structures.

({\it i}) {\it Solenoids (}$\phi \ll 1$, $\theta \ll 1$): We start with the
solenoidal fiber with $\phi \ll \theta \ll 1$. Then the pitch angle is small
(cf. Eq. \ref{alpha2}) and each loop of the solenoid resembles nearly a
circle. The small variations in $\phi $ will add up to an effective
deviation $\Delta \zeta $ from the original orientation of the fiber per
turn of the helix. If one has $n$ monomers per turn it can be shown that $%
\left\langle \Delta \zeta ^{2}\right\rangle =n\sigma _{\phi }^{2}$ with $%
\sigma _{\phi }$ the width of the $\phi $-distribution. $\Delta \zeta $ is
Gaussian with a width $\sigma _{\zeta }=\sqrt{n}\sigma _{\phi }$. With each
turn the middle axis of the solenoid proceeds by a length $d$ where $d$ is
given by Eq. \ref{d}. We can interpret the middle axis of the solenoid as a
new effective chain with bond length $d$, and calculate the average of the
scalar product of an arbitrary pair of successive bond vectors ${\bf a}_{i}$
and ${\bf a}_{i+1}$ of this new effective chain: 
\begin{equation}
\left\langle {\bf a}_{i}{\bf a}_{i+1}\right\rangle =\frac{d^{2}}{\sqrt{2\pi }%
\sigma _{\zeta }}\int d\Delta \zeta \cos \left( \Delta \zeta \right) \exp
\left( -\frac{\Delta \zeta ^{2}}{2\sigma _{\zeta }^{2}}\right) \simeq
d^{2}\left( 1-\frac{\sigma _{\zeta }^{2}}{2}\right) \,,
\end{equation}
the approximation holding for $\sigma _{\zeta }\ll 1$. It follows that the
end-to-end distance of the chain, assuming that the solenoid has $M$ turns
(corresponding to a fiber of $N=2\pi M/\theta $ nucleosomes), is 
\begin{equation}
\left\langle L^{2}\right\rangle
=\sum\limits_{n=1}^{M}\sum\limits_{m=1}^{M}\left\langle {\bf a}_{n}{\bf a}%
_{m}\right\rangle \simeq \frac{4d^{2}M}{\sigma _{\zeta }^{2}}  \label{ee}
\end{equation}
The persistence length $l_{P}$ of the fiber follows from $l_{P}=\left\langle
L^{2}\right\rangle /\left( dM\right) $: 
\begin{equation}
l_{P}\simeq \frac{4\phi }{\theta \sigma _{\phi }^{2}}b  \label{lp}
\end{equation}
where we have made use of the relations $d\simeq \left( 2\pi \phi /\theta
^{2}\right) b$ (cf. Eq. \ref{d}) and $\sigma _{\zeta }^{2}=2\pi \sigma
_{\phi }^{2}/\theta $. These results must be modified for solenoids with
larger pitch angle $\gamma $ ($\theta \ll \phi \ll 1$), where -- since only
the component $\Delta \phi $cos$\gamma $ of a variation $\Delta \phi $ in
the rotational angle $\phi $ leads to a change in the direction of the fiber
(the component $\Delta \phi $sin$\gamma $ leads to a twist) -- one has to
replace $\sigma _{\phi }$ by $\sigma _{\phi }\cos \gamma $. The resulting
persistence length is given by $l_{P}\simeq 4b/\left( \sigma _{\phi
}^{2}\cos ^{2}\gamma \right) \simeq 4b\phi ^{2}/\left( \sigma _{\phi
}^{2}\theta ^{2}\right) $.

({\it ii}) {\it Fiber with crossed linkers} ($\phi \ll 1$, $\pi -\theta \ll
1 $): This case can be calculated analogously. The number of monomers per
''turn'' is given by $n=\pi /\left( \pi -\theta \right) $ (see above) so
that $\sigma _{\zeta }=\sqrt{\pi /\left( \pi -\theta \right) }\sigma _{\phi
} $. Furthermore, the bond length of the new effective chain is $d\simeq \pi
\phi b/4$ (cf. Eq. \ref{d2}). From Eq. \ref{ee} it follows that a fiber of $%
N=nM$ monomers has the mean-squared end-to-end-distance 
\begin{equation}
\left\langle L^{2}\right\rangle \simeq \frac{\pi }{4}\frac{\phi ^{2}\left(
\pi -\theta \right) }{\sigma _{\phi }^{2}}b^{2}M  \label{L4}
\end{equation}
and a persistence length 
\begin{equation}
l_{P}\simeq \frac{\phi \left( \pi -\theta \right) }{\sigma _{\phi }^{2}}b
\label{lp2}
\end{equation}
Now consider typical values for chromatin: $\phi =36{{}^{\circ }}$, $b=20nm$%
, and $\theta \approx 145{{}^{\circ }}$ at $80mM$, $\theta \approx 135{%
{}^{\circ }}$ at $15mM$ and $\theta \approx 95{{}^{\circ }}$ at $5mM$
(Bednar et al., 1998). Assume that the histones are located at equidistant
positions but with small variations, typically $\pm 1bp$, i.e., $\sigma
_{\phi }\approx 36{{}^{\circ }}$; then we find $l_{P}\approx 20nm$ at $80mM$%
, $l_{P}\approx 25nm$ at $15mM$ and $l_{P}\approx 47nm$ at $5mM$.

({\it iii}) {\it Twisted zig-zag fiber}: Finally, we consider zig-zag
structures, first the case where the angle $\phi $ fluctuates around the
mean value $\pi $ (planar zig-zag). (With no fluctuations in $\phi $, the
zig-zag structures simply represent a perfectly flat ribbon.) Assume first
that one bond is slightly rotated by $\Delta \varphi \ll 1$. As a
consequence the ribbon is deflected by an angle $\Delta \zeta _{1}\simeq
\sin \left( \theta /2\right) \Delta \varphi $; furthermore the orientation
of the plane defined by the ribbon rotates by an angle $\Delta \zeta
_{2}\simeq \cos \left( \theta /2\right) \Delta \varphi $. A long ribbon-like
zig-zag structure with small fluctuations of the $\phi $-angle shows
individual configurations typical of a polymer with an anisotropic bending
rigidity (Nyrkova et al., 1996); such a polymer has a plane of main
flexibility, being highly rigid in the direction perpendicular to this
plane. The anisotropy leads to two persistence lengths: an {\it in-plane
persistence length} $l_{1}$ which is associated with the deflection of the
ribbon within the plane of main flexibility; and an {\it out-of-plane
persistence length} $l_{2}$, the typical polymer length that is needed to
''forget'' the orientation of the plane of main flexibility. $l_{1}$ follows
from the number of monomers $n_{1}$ that is needed on average to forget the
original orientation of the axis of the ribbon, $\Delta \zeta
_{1}^{2}n_{1}=4 $ (the numerical value is chosen so that $l_{1}$ is
compatible with the definition of the Kuhn statistical length). Thus 
\begin{equation}
l_{1}=b\cos \left( \theta /2\right) n_{1}\simeq \frac{4\cos \left( \theta
/2\right) }{\sin ^{2}\left( \theta /2\right) }\frac{b}{\sigma _{\phi }^{2}}
\label{lp3}
\end{equation}
Similarly, $l_{2}$ follows from $\left( 2\pi \right) ^{2}=\Delta \zeta
_{2}^{2}n_{2}$: 
\begin{equation}
l_{2}\simeq b\cos \left( \theta /2\right) n_{2}\simeq \frac{4}{\cos \left(
\theta /2\right) }\frac{b}{\sigma _{\phi }^{2}}  \label{lp4}
\end{equation}
We consider the two limiting cases. ({\it i}) $\theta =0$: here $%
l_{1}=\infty $ and $l_{2}\simeq 4b/\sigma _{\phi }^{2}$. The configuration
of the chain is that of a straight line. Variations in $\Delta \varphi $ do
not affect the positions of the monomers. ({\it ii}) $\theta \rightarrow \pi 
$: By reaching this limit the chain collapses into a configuration where it
just goes back and forth between two monomer positions. Indeed, we find from
the above equations that $l_{1}\rightarrow 0$ and $l_{2}\rightarrow \infty $.

For a twisted zig-zag structure with $\phi =\pi -\delta $ with $\delta \ll 1$
there is an inherent orientational persistence length that follows from the
twist of the fiber. This leads to a length $\bar{l}_{2}\simeq b\cos \left(
\theta /2\right) \bar{n}_{2}$ where $\bar{n}_{2}=2\pi /\delta $ denotes the
number of monomers per turn. Apparently the inherent twist competes with the
randomly introduced one, and the out-of-plane persistence length $l_{2}$
(cf. Eq. \ref{lp4}) has to be replaced by $\bar{l}_{2}$ if $l_{2}\lesssim 
\bar{l}_{2}$. The role of variations in the linker length in the case of a
twisted zig-zag structure was simulated by Woodcock et al. (cf. Fig. 3 in
Woodcock et al., 1993). They chose the case $\theta =120{{}^{\circ }}$ and $%
\delta =360{{}^{\circ }}/13\simeq 0.48$. Using Eqs. \ref{lp3} and \ref{lp4}
we find $l_{1}\simeq 2.7b/\sigma _{\phi }^{2}$, $l_{2}\simeq 8.0b/\sigma
_{\phi }^{2}$ and $\bar{l}_{2}=6.5b$. It follows from our formulas that the
persistence lengths $l_{1}$ and $l_{2}$ decay rapidly with $\sigma _{\phi }$%
, a trend that can also be seen in the displayed configuration in Fig. 3 of
Woodcock et al. If we choose, for instance, $\sigma _{\phi }=1/2$ we find $%
l_{p}\simeq 11b$, a persistence similar to that of the fiber displayed in
their Fig. 3b. If we double $\sigma _{\phi }$, i.e., $\sigma _{\phi }=1$, we
find $l_{1}\simeq 3b$ so that there is no longer a well-defined fiber; a
similarly disordered fiber is displayed in their Fig. 3d. A closer
comparison between our theoretical results and the disordered fibers shown
in Woodcock et al. is not possible since in the case of the ''simulated''
fibers a discontinuous distribution of the values of $\phi $ was chosen,
thereby varying the number of base pairs per linker.\bigskip

\begin{center}
{\bf APPENDIX\ C:\ PERSISTENCE\ LENGTHS}
\end{center}

We calculate here the effect of linker {\it flexibility} on the persistence
length of the two-angle fiber. We first calculate the zig-zag-structure
where one has two different persistence lengths, the persistence length $%
l_{P}^{\left( in\right) }$ for bending within the plane of the fiber, and
the length $l_{P}^{\left( out\right) }$ for bending out of the plane.

({\it i}) {\it bending in the plane of the fiber:} Assume that the ribbon is
bent within its plane with a large radius $R$ of curvature so that $R\gg b$.
The linkers are bent but not twisted in this case. Up to corrections of
order $\left( \bar{b}/R\right) ^{2}$ the shape of each linker (i.e. its
deviation from a straight line) is given by $u\left( x\right) =-\varepsilon
x^{2}/\bar{b}+\varepsilon x$. This function fulfills the appropriate
boundary conditions $u\left( 0\right) =u\left( \bar{b}\right) =0$ and $%
u^{\prime }\left( 0\right) =-u^{\prime }\left( \bar{b}\right) =\varepsilon $%
. This leads to the following bending energy per linker: $E=\left( \kappa 
\bar{b}/2R^{2}\right) \cos ^{2}\left( \theta /2\right) $. In the
longitudinal direction of the fiber this corresponds to the bending of a
piece of the length $\bar{b}\cos \left( \theta /2\right) $. Thus 
\begin{equation}
A_{fiber}^{\left( in\right) }=A_{\text{DNA}}\cos \left( \theta /2\right)
\label{lp5}
\end{equation}

({\it ii}) {\it bending perpendicular to the plane of the fiber: }This
bending is accomplished by a combination of twist and bending of the
linkers. Consider the two cases separately. If there is only twist allowed ($%
\kappa \rightarrow \infty $), then each linker has to be twisted by an angle 
$\bar{b}\cot \left( \theta /2\right) /R$ which leads to a twisting energy $%
E=C\cot ^{2}\left( \theta /2\right) \bar{b}/2R^{2}$ and then in turn to the
persistence length $A_{twist}^{\left( out\right) }=\left( C/k_{B}T\right)
\cos \left( \theta /2\right) /\sin ^{2}\left( \theta /2\right) $. Now
consider the case without twisting ($C\rightarrow \infty $) but with bending
of the linker only. If one bends a linker out of the plane of the fiber with
a radius of curvature $R$, it can be shown that as a result the zig-zag is
deflected by an angle $\bar{b}\cos \left( \theta /2\right) /R$. If each
linker is bent in such a way, the zig-zag fiber is bent out of its plane
with an overall curvature of $1/R$. The bending energy per linker is $%
E/k_{B}T=A_{\text{DNA}}\bar{b}/2R^{2}$, leading to a persistence length $%
A_{bend}^{\left( out\right) }=A_{\text{DNA}}/\cos \left( \theta /2\right) $.
By putting the two deformation modes ''in series'' we find the overall
persistence length for bending the zig-zag out of the plane: 
\begin{equation}
A_{fiber}^{\left( out\right) }\simeq \left( 1/A_{twist}^{\left( out\right)
}1+1/A_{bend}^{\left( out\right) }\right) ^{-1}=\frac{A_{\text{DNA}}}{\cos
\left( \theta /2\right) }\frac{1}{1+\frac{\kappa }{C}\tan ^{2}\left( \theta
/2\right) }  \label{lp6}
\end{equation}
For small angles of $\theta $, the bending contribution dominates and $%
A_{fiber}^{\left( out\right) }\rightarrow A_{\text{DNA}}$ for $\theta
\rightarrow 0$ (naked DNA). On the other hand, a very dense zig-zag with a
value of $\theta $ close to $\pi $ is bent by the twisting of the linkers,
leading to a very short persistence length $A_{fiber}^{\left( out\right)
}\simeq \left( C/k_{B}T\right) \cos \left( \theta /2\right) $.
Interestingly, for DNA where $\kappa \approx C$ one finds from Eq. \ref{lp6} 
$A_{fiber}^{\left( out\right) }\approx A_{\text{DNA}}\cos \left( \theta
/2\right) $ over the whole range of $\theta $-values. Thus, in this case $%
A_{fiber}^{\left( in\right) }\approx A_{fiber}^{\left( out\right) }$.

We turn now to fibers with crossed linkers ($\phi $ small, $\theta $ large).
If we bend such a fiber within a given plane, then an inhomogeneous
deformation pattern result where some of the linkers are oriented (nearly)
parallel to the fiber while others are perpendicular. The first class of
linkers will be bent, the second will be mostly twisted. The effective angle
is now $\pi -\phi \cot \left( \theta /2\right) $ instead of $\theta $ (this
follows from $L$ in Table I with $N=2$). Since $C\approx \kappa $ the
contribution to the elastic energy is approximately the same for all the
linkers, leading to a persistence length 
\begin{equation}
A_{fiber}\approx A_{\text{DNA}}\cos \left( \frac{\pi }{2}-\frac{\phi }{2}%
\cot \left( \theta /2\right) \right) \simeq A_{\text{DNA}}\frac{\phi }{2}%
\cot \left( \theta /2\right)  \label{lp7}
\end{equation}
Using the $\theta $-values given by Bednar et al., $\phi \approx 36{%
{}^{\circ }}$ (and $A_{\text{DNA}}\approx 50nm$) we find $A_{fiber}\approx
14nm$ for $\theta =95{{}^{\circ }}$ (the value at 5mM monovalent salt), $%
A_{fiber}\approx 6nm$ for $\theta =135{{}^{\circ }}$ (15mM) and $%
A_{fiber}\approx 5nm$ for $\theta =145{{}^{\circ }}$ (80mM). These values
are smaller than the diameter of the fiber so it is reasonable to assume $%
A_{fiber}\approx 30nm$ for the persistence length of the 30-nm fiber
(roughly independent of the salt concentration).

\begin{center}
{\bf APPENDIX\ D:\ COMPARISON\ WITH\ COMPUTER\ SIMULATIONS}
\end{center}

Katritch et al. performed Monte-Carlo simulations of the two-angle model
with flexible linkers (Katritch et al., 2000). Here we compare their results
with our theoretical predictions. We first base our analysis on the results
for the zig-zag case, Eq. \ref{force2}, for reasons given below. Including
the entropic contribution we find the following force law: 
\begin{equation}
f=\left\{ 
\begin{array}{ll}
\frac{3k_{B}T}{l_{P}\cos \left( \theta /2\right) }x & \mbox{for}\;x\ll \cos
\left( \theta /2\right) \\ 
\frac{12}{\sin ^{2}\left( \theta /2\right) }\frac{\kappa }{\bar{b}^{2}}\left[
x-\cos \left( \theta /2\right) \right] +\frac{3k_{B}T}{l_{P}} & \mbox{for}%
\;x\gg \cos \left( \theta /2\right)
\end{array}
\right.  \label{force4}
\end{equation}
To allow a better comparison with the diagrams in Fig. 3 of Katritch et al.
we present in Eq. \ref{force4} the force as a function of the relative
extension $x=L/\left( \bar{b}N\right) $. While most of the data of Katritch
et al. were obtained assuming a (quenched) set of random values of the
rotational setting of the nucleosomes, we believe that the {\it qualitative}
dependence of $f$ on $\theta $ and $\bar{b}$ should be unaffected by this
assumption. Figure 3(a) in Katritch et al. shows the dependence of the
force-extension profile on the entry-exit angle $W=\pi -\theta $. It can be
seen that the initial slope (entropy regime) decreases with $W$ (increases
with $\theta $) in accordance with Eq. \ref{force4}. The behavior at larger
forces shows the opposite dependence, as is also predicted by Eq. \ref
{force4}. Finally, the crossover is shifted with increasing $W$ to larger
values; this again is in accordance with Eq. \ref{force4}. Figure 3(b) in
Katritch et al. depicts the dependence on the linker length $\bar{b}$; here
the data show no indication of a dependence of the initial slope on $\bar{b}$%
, and similarly for the value of the crossover, all in accordance with Eq. 
\ref{force4}. Then for the second regime they find an increasing slope with
decreasing linker length, also in accord the above theory.

Fig. 3(c) of Katritch et al. shows a comparison between random rotational
settings of the nucleosomes and non-random settings. The zig-zag case, $\phi
=\pi $, has a slightly smaller initial slope than the random case and a
slightly larger slope in the second linear regime, but follows always quite
closely the case of random settings. This justifies the above comparison
between Eq. \ref{force4} --- based on the zig-zag case --- and the computer
simulations using a random setting. Figure 3(c) of Katritch et al. allows
now also a direct {\it quantitative} comparison. We predict from Eq. \ref
{force4}, for the values $\theta =2.27$ and $\bar{b}=40bp$ (used in Katritch
et al.), that $f\simeq 0.95x$ for $x\lesssim 0.42$ and $f\simeq 14.9\left(
x-0.42\right) +0.4$ for $x\gtrsim 0.42$ (force in $pN$), in good agreement
with their data points. (For $x\gtrsim 0.6$ the datapoints indicate an
increasing slope, a result of nonlinear effects not taken into account in
our theory.)

Finally, Katritch et al. also provide data points for the case $\phi =0.35$
and $\theta =2.27$. We find from Eq. \ref{fregimes} that one has $f\simeq
4.9x$ for $x\lesssim 0.1$ and $f\approx 8.6\left( x-0.1\right) +0.5$ for $%
x\gtrsim 0.1$. (Here $x\equiv L/\bar{b}N$, and $x_{0}\equiv L_{0}/\bar{b}%
N=\left( \phi /2\right) \cot \left( \theta /2\right) \simeq 0.1$.) This
result overestimates the entropic contribution as can be seen by comparison
with the $Tw=20{{}^{\circ }}$--curve in Fig. 3(c) of Katritch et al. (data
points are missing for $x<0.2$ but the force at $x=0.2$ is smaller than $1pN$%
). The reason is probably the underestimation of the persistence length of
the fiber $l_{P}$, assumed here to be of the order of the fiber thickness.
The real value could be larger since this set of angles corresponds to an
extremely dense fiber where excluded volume effects become rather important.
The simulation data for this set of angles is not found to be in good
agreement with the experimental force-extension characteristics. This might
be attributed to the small value of the entry-exit angle ($130{{}^{\circ }}$
instead of $85{{}^{\circ }}$ as suggested by the electron cryomicrographs,
Bednar et al., 1998); if we use $\theta =95{{}^{\circ }}$ and $\phi =36{%
^{\circ }}$ (as above) we find $f\simeq 1.4x$ for $x\lesssim 0.3$ and $%
f\approx 4.7\left( x-0.3\right) +0.2$ for $x\gtrsim 0.3$ which is closer to
the experimental curve.

\newpage

\begin{center}
{\bf REFERENCES}
\end{center}

\begin{quote}
Bednar, J., R. A. Horowitz, J. Dubochet, and C. L. Woodcock. 1995. Chromatin
conformation and salt-induced compaction -- 3-dimensional structural
information from cryoelectron microscopy. {\it J. Cell. Biol.} 131:1365-1376.

Bednar, J., R. A. Horowitz, S. A. Grigoryev, L. M. Carruthers, J. C. Hansen,
A. J. Koster, and C. L. Woodcock. 1998. Nucleosomes, linker DNA, and linker
histone form a unique structural motif that directs the higher-order folding
and compaction of chromatin. {\it Proc. Natl. Acad. Sci. USA}.
95:14173-14178.

Butler, P. J. G., J. O. Thomas. 1998. Dinucleosomes show compaction by ionic
strength, consistent with bending of linker DNA. {\it J. Mol. Biol. }%
281:401-407.

Cluzel P., A. Lebrun, C. Heller, R. Lavery, J. L. Viovy, D. Chatenay, and F.
Caron. 1996. DNA: An extensible molecule. {\it Science} 271:792-794.

Crothers, D. M., J. Drak, J. D. Kahn, and S. D. Levene. 1992. DNA bending,
flexibility, and helical repeat by cyclization kinetics. {\it Meth.
Enzymology} 212:3-29.

Cui, Y., and C. Bustamante. 2000. Pulling a single chromatin fiber reveals
the forces that maintain its higher-order structure. {\it Proc. Natl. Acad.
Sci. USA. }97: 127-132.

Doi, M., and S. F. Edwards. 1986. The theory of polymer dynamics. Clarendon
Press, Oxford.

de Gennes, P.-G. 1979. Scaling concepts in polymer physics. Cornell
University Press, Ithaca.

Finch, J. T., and A. Klug. 1976. Solenoidal model for superstructure of
chromatin. {\it Proc. Natl. Acad. Sci. USA. }73:1897-1901.

Fraden, S., and R. D. Kamien. 2000. Self-assembly in vivo. {\it Biophys. J. }%
78:2189-2190.

Hagerman, P. J. 1988. Flexibility of DNA. {\it Annu Rev Biophys Biophys Chem.%
} 17:265-286.

Horowitz, R. A., D. A. Agard, J. W. Sedat, and C. L. Woodcock. 1994. The
three-dimensional architecture of chromatin in situ: electron tomography
reveals fibers composed of a continuously variable zig-zag nucleosomal
ribbon. {\it J. Cell. Biol. }125:1-10.

Katritch, V., C. Bustamante, and W. K. Olson. 2000. Pulling chromatin
fibers: computer simulations of the direct physical micromanipulation. {\it %
J. Mol. Biol. }295:29-40.

Khrapunov, S. N., A. I. Dragan, A. V. Sivolob, and A. M. Zagariya. 1997.
Mechanisms of stabilizing nucleosome structure. Study of dissociation of
histone octamer from DNA. {\it Biochim. Biophys. Acta.} 1351:213-222.

Klenin, K. V., A. V. Vologodskii, V. V. Anshelevich, V. Y. Klishko, A. M.
Dykhne, and M. D. Frank-Kamenetskii. 1989. Variance of writhe for wormlike
DNA rings with excluded volume. {\it J. Biomol. Struct. Dyn.} 6:707-714.

Leuba, S. H., G. Yang, C. Robert, B. Samori, K. van Holde, J. Zlatanovam,
and C. Bustamante. 1994. Three-dimensional structure of extended chromatin
fibers as revealed by tapping-mode scanning force microscopy. {\it Proc.
Natl. Acad. Sci. USA. }91:11621-11625.

Livolant, F., and A. Leforestier. 2000. Chiral discotic columnar germs of
nucleosome core particles. {\it Biophysical Journal. }78:2716-2729.

Luger, K., A. W. M\"{a}der, R. K. Richmond, D. F. Sargent, and T. J.
Richmond. 1997. Crystal structure of the nucleosome core particle at 2.8A
resolution. {\it Nature. }389:251-260.

Marko, J. F., and E. D. Siggia. 1997. Driving proteins off DNA using applied
tension. {\it Biophysical Journal}. 73:2173-2178.

Marko, J. F. 1998. DNA under high tension: Overstretching, undertwisting,
and relaxation dynamics. {\it Phys. Rev. E.} 57:2134-2149.

Nyrkova, I. A., A. N. Semenov, J.-F. Joanny, and A. R. Khokhlov. 1996.
Highly anisotropic rigidity of ''ribbon-like'' polymers: I. Chain
configuration in dilute solutions. {\it J. Phys. II France}. 6:1411-1428.

Plewa, J. S., and T. A. Witten. 2000. Conserved linking in single- and
double-stranded polymers. {\it J. Chem. Phys.} 112:10042-10048.

Polach, K. J., and J. Widom. 1996. A model for the cooperative binding of
eukaryotic regulatory proteins to nucleosomal target sites. {\it J. Mol.
Biol.} 254:800-812.

Raspaud, E., I. Chaperon, A. Leforestier, and F. Livolant. Spermine-induced
aggregation of DNA, nucleosome, and chromatin. {\it Biophys. J.}
77:1547-1555.

Schiessel, H., J. Widom, R. F. Bruinsma, and W. M. Gelbart. Polymer
reptation and nucleosome repositioning. {\it preprint}

Schlick, T. 1995. Modeling superhelical DNA -- recent analytical and dynamic
approaches. {\it Curr. Opin. Struc. Biol.} 5:245-262.

Smith, S. B., Y. Cui, and C. Bustamante. 1996. Overstretching B-DNA: The
elastic response of individual double-stranded and single-stranded DNA
molecules. {\it Science}. 271:795-799.

Stryer, L. 1995. Biochemistry, 4'th edition. Freeman. pg. 420-421.

Thoma, F., Th. Koller, and A. Klug. Involvement of the histone H1 in the
organization of the nucleosome and of the salt-dependent superstructures of
chromatin. {\it J. Cell. Biol. }83:403-427.

van Holde, K. E. 1989. Chromatin. Springer Verlag, New York.

van Holde, K., and J. Zlatanova. 1995. Chromatin higher order structure:
chasing a mirage? {\it J. Biol. Chem. }93:8373-8376.

van Holde, K., and J. Zlatanova. 1996. What determines the folding of the
chromatin fiber? {\it Proc. Natl. Acad. Sci. }93:10548-10555.

Widom, J. and A. Klug. 1985. Structure of the 300A\ chromatin filament:
X-ray diffraction from oriented samples. {\it Cell. }43:207-213.

Widom, J. 1986. Physicochemical studies of the folding of the 100A\
nucleosome filament into the 300A\ filament. {\it J. Mol. Biol.} 190:411-424.

Widom., J. 1989. Toward a unified model of chromatin folding. {\it Annu.
Rev. Biophys. Biophys. Chem. }18:365-395.

Widom, J. 1992. A relationship between the helical twist of DNA and the
ordered positioning of nucleosomes in all eukaryotic cells. {\it Proc. Natl.
Acad. Sci. USA. }89:1095-1099.

Widom, J. 1998. Structure, dynamics, and function of chromatin in vitro. 
{\it Annu. Rev. Biophys. Biomol. Struct. }27:285-327.

Woodcock, C. L., S. A. Grigoryev, R. A. Horowitz, and N. Whitaker. 1993. A
chromatin folding model that incorporates linker variability generates
fibers resembling native structures. {\it Proc. Natl. Acad. Sci. USA. }%
90:9021-9025.

Yao J., Lowary P. T., and Widom J. 1990. Direct detection of linker DNA
bending in defined-length oligomers of chromatin. {\it Proc. Natl. Acad.
Sci. USA} 87:7603-7607.

\newpage

Table I\bigskip

\begin{tabular}{|c|c|c|c|}
\hline
& $
\begin{array}{l}
\text{solenoid} \\ 
\phi \ll 1\text{, }\theta \ll 1
\end{array}
$ & $
\begin{array}{l}
\text{crossed\thinspace \thinspace linkers} \\ 
\phi \ll 1\text{, }\pi -\theta \ll 1
\end{array}
$ & $
\begin{array}{l}
\text{twisted zig-zag} \\ 
\phi =\pi -\delta \text{ with }\delta \ll 1
\end{array}
$ \\ \hline
$
\begin{array}{l}
\end{array}
R 
\begin{array}{l}
\end{array}
$ & $\frac{\theta }{\phi ^{2}+\theta ^{2}}b$ & $\frac{b}{2\sin \left( \theta
/2\right) }\left( 1-\frac{\phi ^{2}}{4}\cot ^{2}\left( \theta /2\right)
\right) $ & $\frac{b}{2}\sin \left( \frac{\theta }{2}\right) \left( 1+\frac{%
\delta ^{2}}{4}\right) $ \\ \hline
$
\begin{array}{l}
\end{array}
L 
\begin{array}{l}
\end{array}
$ & $\frac{\phi bN}{\sqrt{\phi ^{2}+\theta ^{2}}}$ & $\frac{N\phi b}{2}\cot
\left( \theta /2\right) $ & $b\cos \left( \frac{\theta }{2}\right) N\left( 1-%
\frac{\delta ^{2}}{8}\tan ^{2}\left( \frac{\theta }{2}\right) \right) $ \\ 
\hline
$
\begin{array}{l}
\end{array}
\lambda 
\begin{array}{l}
\end{array}
$ & $\frac{\sqrt{\phi ^{2}+\theta ^{2}}}{\phi b}$ & $\frac{4}{b\phi \left(
\pi -\theta \right) }$ & $\frac{1+\delta ^{2}\tan ^{2}\left( \theta
/2\right) /8}{b\cos \left( \theta /2\right) }$ \\ \hline
$
\begin{array}{l}
\end{array}
\rho 
\begin{array}{l}
\end{array}
$ & $\frac{\left( \phi ^{2}+\theta ^{2}\right) ^{5/2}}{\pi \phi \theta
^{2}b^{3}}$ & $\frac{16}{\pi \phi \left( \pi -\theta \right) b^{3}}$ & $%
\frac{4}{\pi }\frac{1+\delta ^{2}\left( \tan ^{2}\left( \theta /2\right)
/4-1\right) /2}{b^{3}\cos \left( \theta /2\right) \sin ^{2}\left( \theta
/2\right) }$ \\ \hline
\end{tabular}

\smallskip \bigskip

\bigskip {\it Table caption:}

Table I: Geometrical properties of the two-angle fiber for the three
limiting cases: solenoid, fiber with crossed linkers and twisted zig-zag
fiber. Displayed are the fiber radius $R$, the length $L$ of a $\left(
N+1\right) $-mer, the line density $\lambda =N/L$, and the 3D density $\rho
=\lambda /\pi R^{2}$.\newpage {\it Figure captions}:

Figure 1: Schematic representation of the nucleosome. Eight core histones
aggregate into the histone octamer that acts as a cylindrical spool around
which the DNA is wound in 1-and-$3/4$ turns. The linker histone is also
depicted that acts at the entry-exit point of the DNA. The entry-exit angle $%
\pi -\theta $ of the linker DNA is one of the angles defining the two-angle
model.

Figure 2: The two competing models for the 30-fiber: (a) the solenoid model
and (b) the crossed linker model (see text).

Figure 3: Fraction of a two-angle fiber containing four nucleosomes (it is a
part of structure ''11'' in Fig. 4). The two angles are depicted, the
deflection angle $\theta $ and the rotational angle $\phi $, together with
the ''nucleosome diameter'' $2a$ and the ''linker length'' $b$. All four are
considered to be constant throughout the fiber. The arrows denote the
nucleosomal axes, cf. Fig. 1.

Figure 4: Diagram of geometrical states of the two-angle model. Shown are
examples of different configurations and their location in the $\left(
\theta ,\phi \right) $-space (arrows). The dashed and dotted curves depict
the boundaries to the $\left( \theta ,\phi \right) $-values that are
forbidden due to excluded volume interaction, one regime (large $\theta $%
-values) due to ''short-range'' interaction, the other (small $\phi $%
-values) due to ''long-range'' interaction (see text for details).

Figure 5: Stretching of a zig-zag chain. (a) The unperturbed chain ($F=0$)
has a total length $L_{0}$ and straight linkers. (b) The same fiber under
tension $F>0$. The fiber is stretched to an end-to-end distance $L>L_{0}$ by
bending of the linkers. The linkers are bent in such a way that the
entry-exit angles at the individual nucleosomes remain unchanged.

Figure 6: (a) Internucleosomal interaction potential $U$ between nucleosome $%
i$ and $i+2$ as a function of distance $x$. In addition to the elastic
contribution there is an short range attraction for nucleosome at close
contact, $x=2a$. The different curves correspond to different values of the
angle $\theta $. Curve ''1'' has the global minimum at large $x$ (swollen
state ''S'') whereas curve ''3'' has the minimum for nucleosomes in close
contact (condensed state ''C''). Curve ''2'' corresponds to the transition
point. Also depicted is the common tangent for curve ''2''. Its slope
corresponds to the critical stretching force $f_{CS}$ at which nucleosomes
are transferred from the condensed to the stretched state. (b)
Force-extension curve of a condensed fiber (say, the fiber with the
interaction potential ''3''). For extensions $L$ with $L_{1}<L<L_{2}$ one
finds a coexistence plateau with the restoring force $f_{CS}$ (see text for
details).
\end{quote}

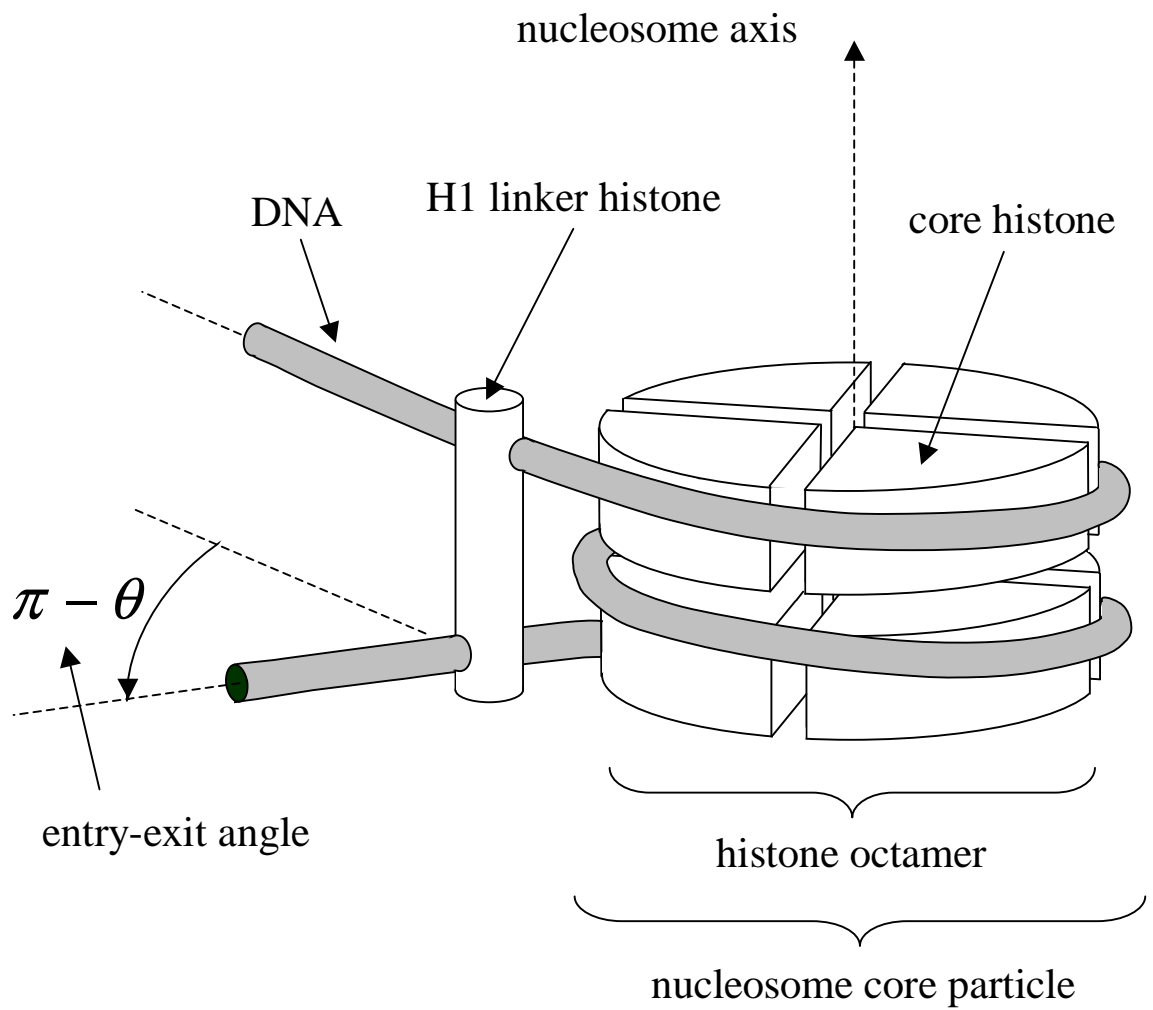
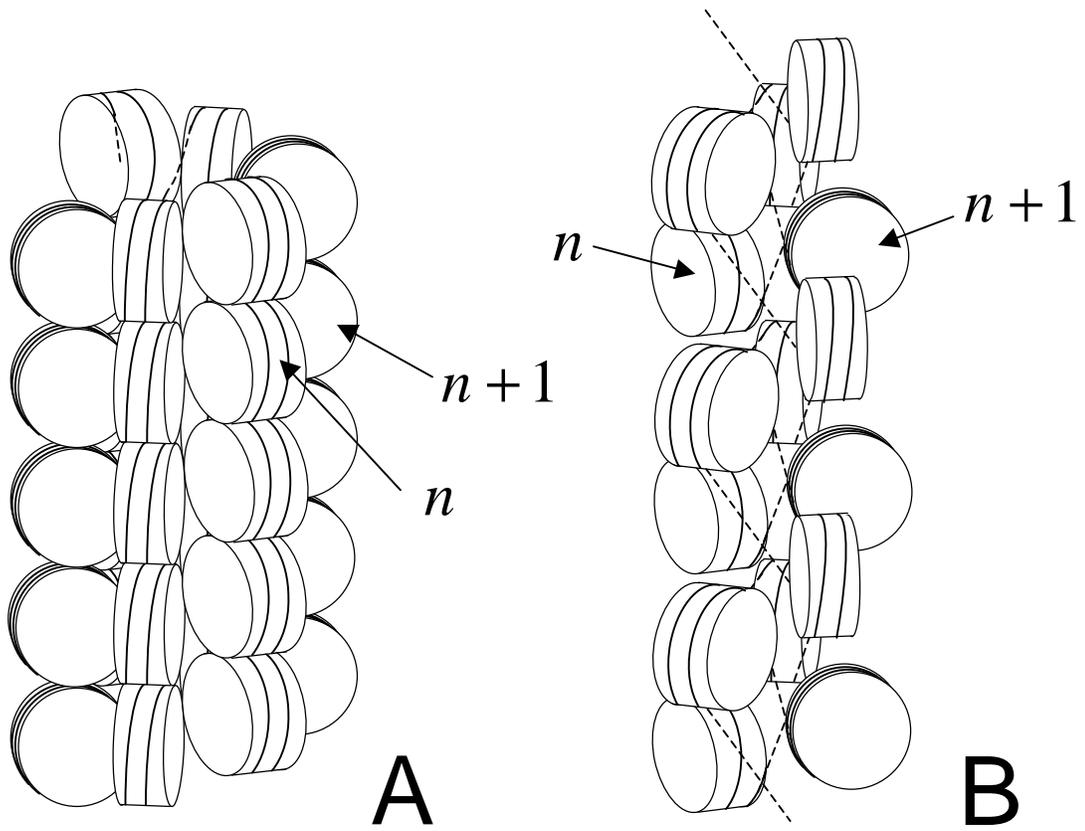
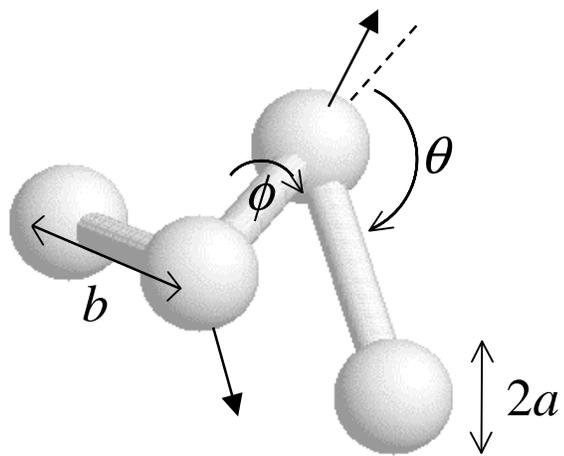
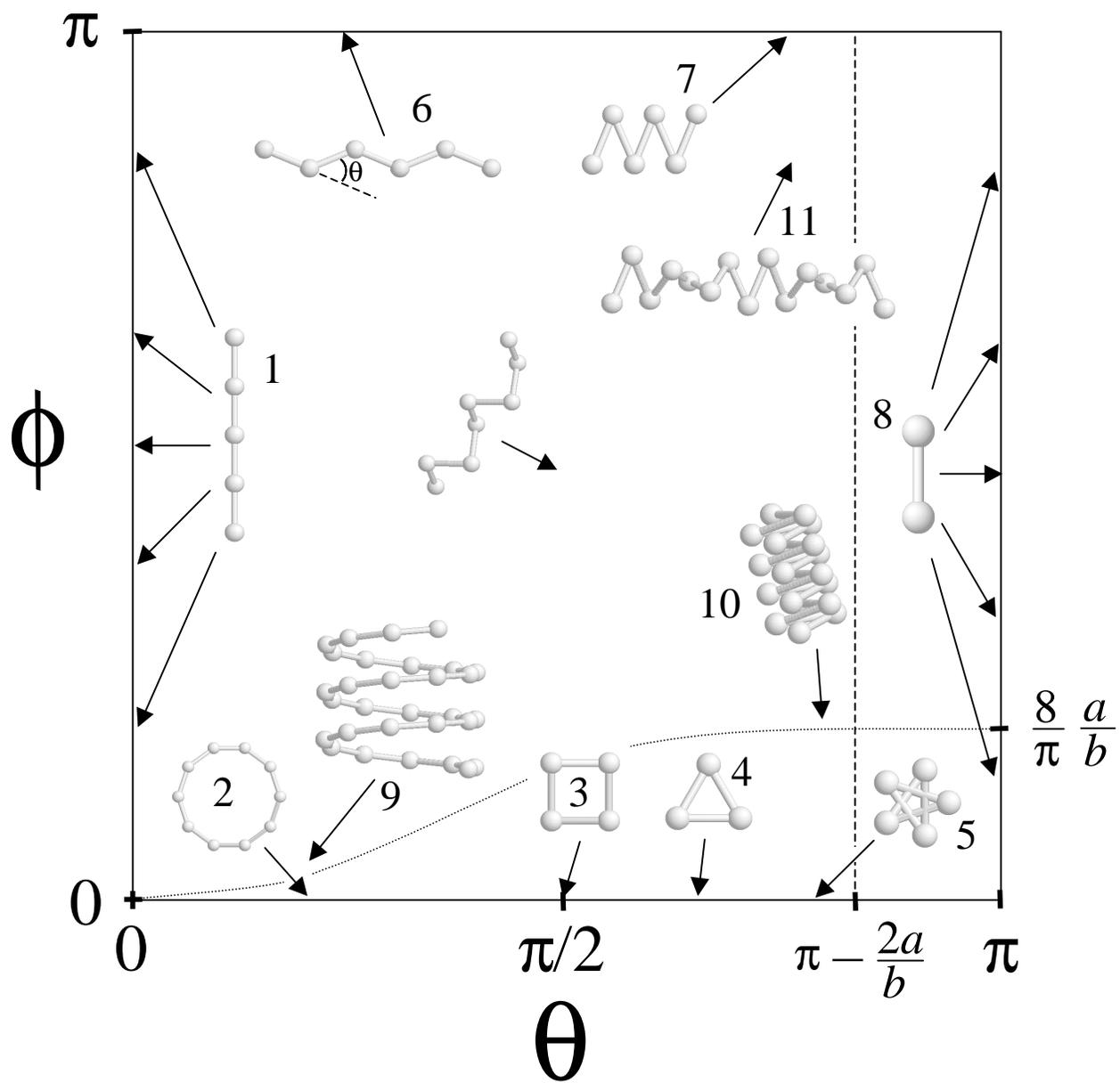
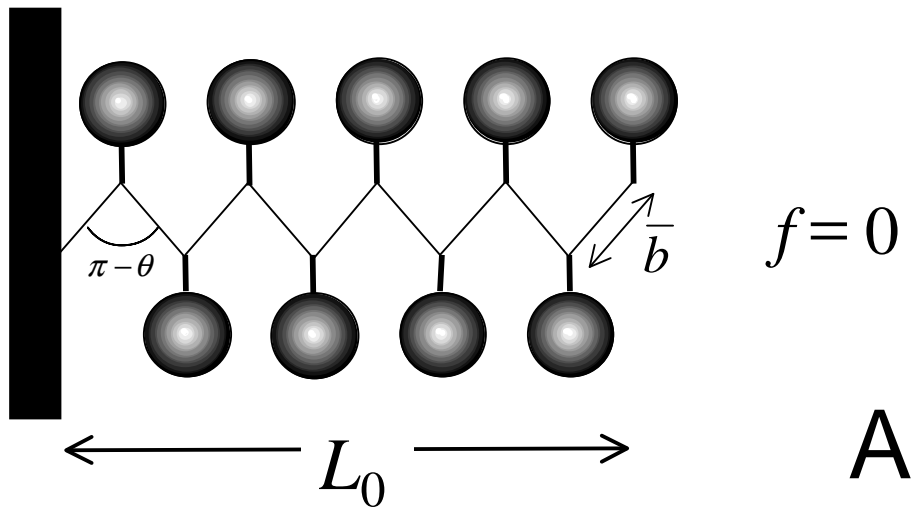
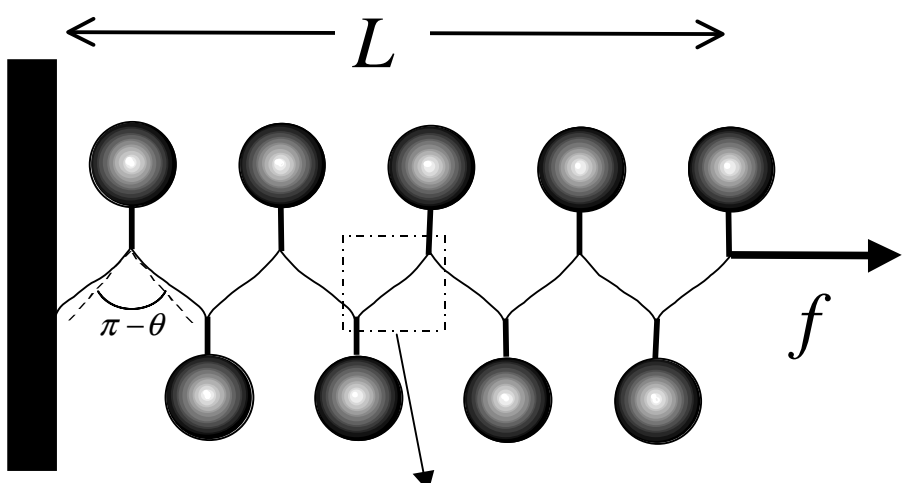
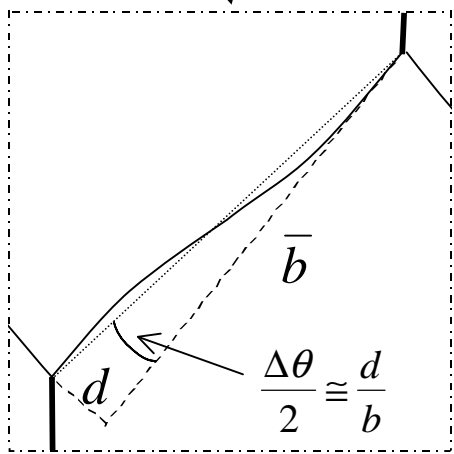
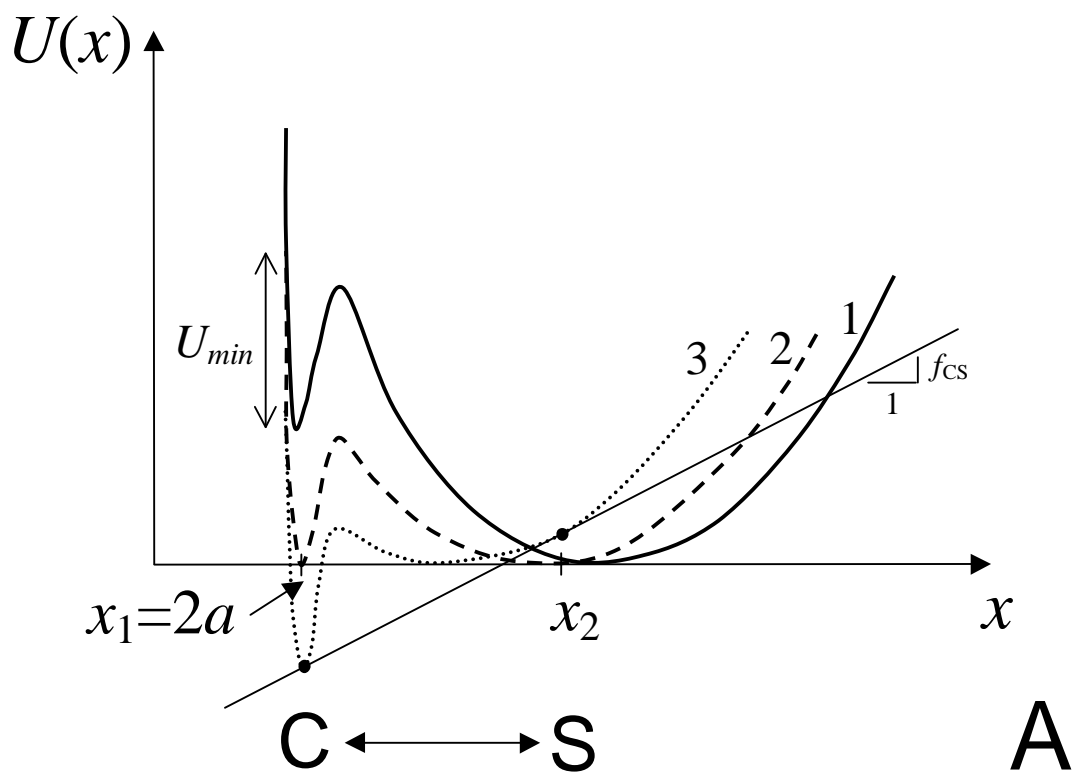
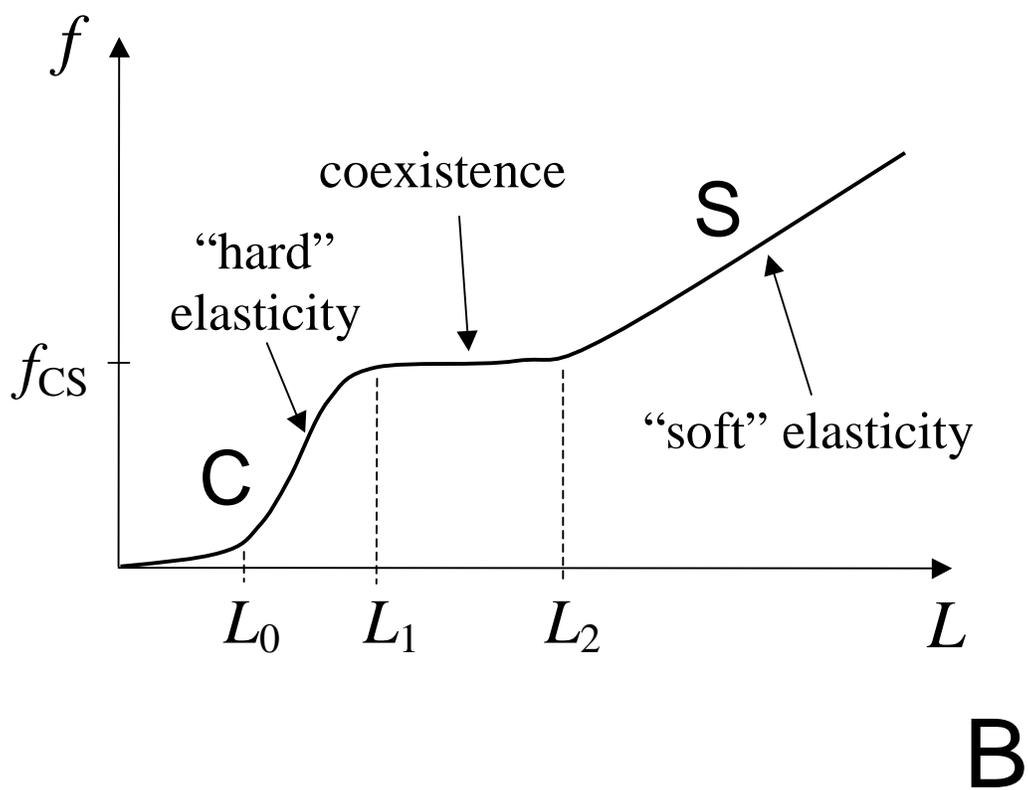
\end{document}